\documentclass[bibyear]{aa}
\usepackage{hyperref}
\usepackage{amsmath,empheq,graphicx,txfonts,amssymb,natbib,epstopdf,grffile,mathtools,lscape,array,color,afterpage}
\usepackage{color}

\def\ls{\mathrel{\hbox{\rlap{\hbox{\lower4pt\hbox{$\sim$}}}\hbox{$<$}}}}
\def\gs{\mathrel{\hbox{\rlap{\hbox{\lower4pt\hbox{$\sim$}}}\hbox{$>$}}}}

\definecolor{light-gray}{gray}{0.65}
\definecolor{light-light-gray}{gray}{0.75}
\definecolor{gray}{gray}{0.4}
\definecolor{dark-blue}{RGB}{0,30,255}
\definecolor{lime-green}{RGB}{114,255,0}
\defcitealias{Pierre2016}{Paper~I}
\defcitealias{Pacaud2016}{Paper~II}
\defcitealias{Giles2016}{Paper~III}
\defcitealias{Democles2015}{D{\'e}mocl{\`e}s, et al. (in prep.)}

  \authorrunning{M.\ Lieu et al.}
  \titlerunning{The XXL Survey IV. $\rm M_{WL}-T_{\rm}$ relation of the XXL-100-GC.}

\begin{document}

\title{The XXL Survey\thanks{Based on observations obtained with XMM-Newton, an ESA sci- ence mission with instruments and contributions directly funded by ESA Member States and NASA. Based on observations made with ESO Telescopes at the La Silla Paranal Observatory under programme 089.A-0666 and LP191.A-0268}
}
\subtitle{ IV. Mass-temperature relation of the bright
  cluster sample}

\author{
  M. Lieu\inst{1}, 
  G. P. Smith\inst{1},   
  P. A. Giles\inst{2}, 
  F. Ziparo\inst{1},
  B. J. Maughan\inst{2}, 
  J. D\'emocl\`es\inst{1},
  F. Pacaud\inst{3},
  M. Pierre\inst{4},
  C. Adami\inst{5},
  Y. M. Bah\'e\inst{6,7},
  N. Clerc\inst{8},
  L. Chiappetti\inst{10},
  D. Eckert\inst{9,10},
  S. Ettori\inst{11,12},
  S. Lavoie\inst{13},
  J.\ P.\ Le~Fevre\inst{14}, 
    I. G. McCarthy\inst{15},
  M. Kilbinger\inst{4},
  T. J.\ Ponman\inst{1}, 
  T. Sadibekova\inst{4}, 
  J. P.\ Willis\inst{7}.}
  \institute{ 
    School of Physics and Astronomy, University of Birmingham, Edgbaston, Birmingham, B15 2TT, England\\
    \email{mlieu@star.sr.bham.ac.uk} \and
    H.\ H.\ Wills Physics Laboratory, University of Bristol, Tyndall Avenue, Bristol, BS8 1TL, England \and
    Argelander Institut f\"ur Astronomie, Universit\"at Bonn, D-53121 Bonn, Germany \and
    Service d'Astrophysique AIM, CEA Saclay, F-91191 Gif sur Yvette, France \and
    Universit\'e Aix Marseille, CNRS, LAM (Laboratoire d'Astrophysique de Marseille) UMR 7326, F-13388, Marseille, France\and
    Max-Planck-Institut f\"ur Astrophysik, Karl-Schwarzschild Str. 1, D-85748 Garching, Germany \and
    Institute of Astronomy, University of Cambridge, Madingley Road, Cambridge CB3 0HA, UK \and
    Max Planck Institut f\"ur Extraterrestrische Physik, Postfach 1312, D-85741 Garching bei M\"unchen, Germany \and
    Department of Astronomy, University of Geneva, ch. d'Ecogia 16, 1290 Versoix, Switzerland \and
    INAF - IASF-Milano, Via E. Bassini 15, I-20133 Milano, Italy \and
    INAF - Osservatorio Astronomico di Bologna, Via Ranzani 1, I-40127 Bologna, Italy\and
    INFN, Sezione di Bologna, viale Berti Pichat 6$\backslash$2, I-40127 Bologna, Italy\and
    Department of Physics and Astronomy, University of Victoria, 3800 Finnerty Road, Victoria, BC, V8P 1A1, Canada\and
    SEDI CEA Saclay, France \and   
    Astrophysics Research Institute, Liverpool John Moores University, IC2, 146 Brownlow Hill, Liverpool L3 5RF, UK\\
  }

\date{\emph{Astronomy \& Astrophysics, status}}
\abstract{The XXL survey is the largest survey carried
  out by \emph{XMM-Newton}. Covering an area of 50 deg$^2$, the
  survey contains $\sim450$ galaxy clusters out to a redshift $\sim$2 and
  to an X-ray flux limit of $\sim5\times10^{-15}{\rm
    erg\,s^{-1}\,cm^{-2}}$. This paper is part of the first release of
  XXL results focussed on the bright cluster sample.}
{We investigate the scaling relation between weak-lensing mass and X-ray temperature for the brightest clusters in XXL. The scaling relation discussed in this article is used to estimate the mass of all 100 clusters in XXL-100-GC.   }
{Based on a subsample of 38 objects that lie within the intersection of the northern XXL field and the publicly available CFHTLenS shear catalog, we derive the weak-lensing mass of each system with careful considerations of the systematics. The clusters lie at  $0.1<z<0.6$ and span a temperature range of $ T\simeq1-5{\rm\ keV}$. We combine our sample with an additional 58 clusters from the literature, increasing the range to $T\simeq1-10{\rm\ keV}$. To date, this is the largest sample of clusters with weak-lensing mass measurements that has been used to study the mass-temperature relation.}
{ The mass--temperature relation fit (M $\propto$ T$^{b}$) to the XXL clusters returns a slope
  $b=1.78^{+0.37}_{-0.32}$ and intrinsic scatter$\sigma_{\ln
    M|T}\simeq0.53$; the scatter is dominated by disturbed clusters. The fit to the combined sample of  96 clusters is in tension with self-similarity, $b=1.67\pm0.12$ and $\sigma_{\ln M|T}\simeq0.41$. }
    { Overall our results demonstrate the feasibility
  of ground-based weak-lensing scaling relation studies down to cool
  systems of $\sim1{\rm keV}$ temperature and highlight that the current data
  and samples are a limit to our statistical precision. As such we are unable to determine whether the validity of hydrostatic equilibrium is a function of halo mass.  An enlarged sample of
  cool systems, deeper weak-lensing data, and robust modelling of the
  selection function will help to explore these issues further.}

\maketitle

\section{Introduction}

Analytical and numerical calculations both predict that the
temperature of the X-ray emitting atmospheres of galaxy groups and
of clusters scales with the mass of their host dark matter halos, with
$M\propto T^{3/2}$ \citep{Kaiser1986,Evrard2002,Borgani2004}.  Testing this so-called self-similar prediction is of
fundamental importance to a broad range of astrophysical and
cosmological problems, including constraining any non-gravitational
physics that affects the gas, and exploring galaxy clusters as probes
of cosmological parameters.

To date, any studies of the mass-temperature relation have employed
X-ray observations to measure both the temperature and the mass of
galaxy groups and clusters. Assuming hydrostatic equilibrium, the self-similar predicted slope value of 1.5 can be derived from the virial theorem. Observational relations, however, generally
steepen from close to the self-similar for hot
systems to a slope of $\sim1.6-1.7$ when cooler systems ($T\ls3{\rm\
  keV}$) are included \citep[see][for recent reviews]{Bohringer2012, Giodini2013}.  These results are subject to several problems,
most prominently that the mass measurements are based on the
assumption that the intracluster gas is in hydrostatic equilibrium
and also that the same data are used for both temperature and mass
measurements, likely introducing a subtle covariance into the analysis.

\begin{figure*}
  \centerline{
    \includegraphics[width=0.5\linewidth, angle=270]{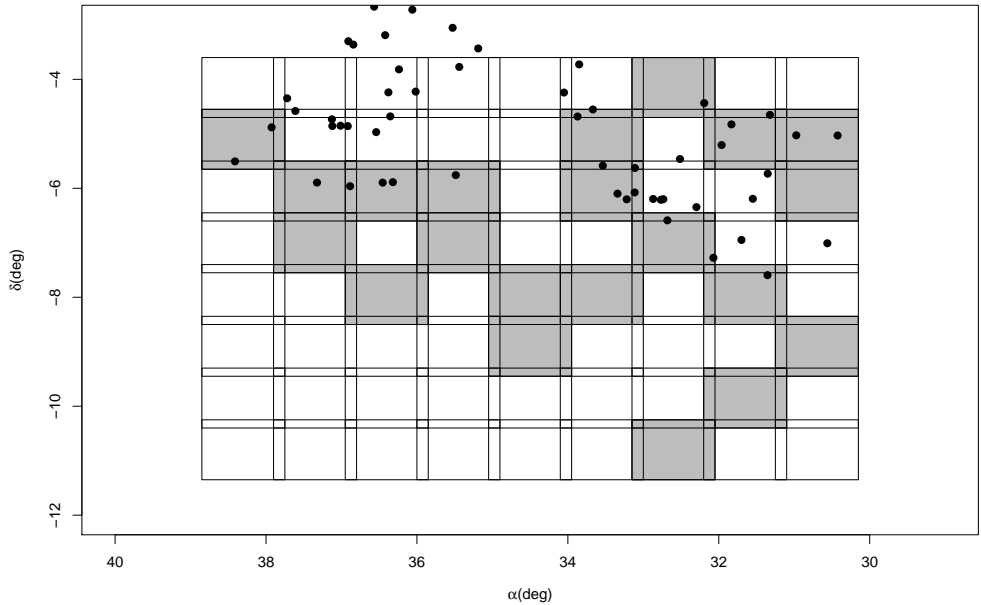}
  }
  \caption{Overlap of XXL-100-GC with the CFHTLenS W1 field. The boxes are individual pointings in CFTHT with
    XXL-North field clusters (filled points). The shaded boxes are pointings that fail the CFHTLenS weak-lensing field selection criteria (See section
    \ref{sec:systematics}). \label{fig:CFHTfield}
  }
\end{figure*}

Independent measurements of mass and temperature, and reliance on
fewer assumptions, help to alleviate these questions.  Gravitational
lensing mass measurements are useful in this regard, and have been
shown to recover the ensemble mass of clusters to reasonably good
accuracy \citep{Becker2011, Bahe2012}, despite
concerns that individual cluster mass measurements may be affected
by halo triaxiality and projection effects \citep[e.g.][]{Corless2007,Meneghetti2010}.  Lensing based studies of the
mass-temperature relation have so far obtained slopes that are
consistent with the self-similar prediction, albeit with large
statistical uncertainties \citep{Smith2005, Bardeau2007, Hoekstra2007, Okabe2010, Jee2011, Mahdavi2013}.  One of the limitations of these studies has been that
they concentrate on relatively hot clusters, $T\gs4{\rm\ keV}$.

Building on the \citet{Leauthaud2010} weak-lensing study of the
mass-luminosity relation of groups in the COSMOS survey, \citet{Kettula2013} recently pushed lensing-based studies of the
mass-temperature relation into the group regime, $T\simeq1-3\ {\rm
  keV}$.  Combining ten groups with complementary measurements of
massive clusters from the literature, they obtained a relation
spanning $T\simeq1-10{\rm\ keV}$, with a slope in good agreement with
the self-similar prediction. This suggests that the assumption of
hydrostatic equilibrium may be less valid in cooler systems than
hotter systems since the discrepancy is only seen at the cool end of the M$_{\rm HSE}$--T relation.
 However, \citet{Connor2014} obtained a slope steeper
than the hydrostatic results using a sample of 15 poor clusters.
Their study was limited to cluster cores within
$r_{2500}$ (i.e.\ the radius at which the mean density of the cluster
is 2500 times the critical density of the universe at the cluster
redshift), in contrast to previous results \citep[e.g][]{Kettula2013} that were derived within
$r_{500}$, indicating that the mass temperature relation may depend on the
cluster centric radius within which the mass is measured.

We present the mass calibration of the XXL bright cluster sample (XXL-100-GC)
based on a new mass-temperature relation that we constrain using the
largest sample used to date for such studies:  96  groups and clusters
spanning X-ray temperatures of $T\simeq1-10{\rm\ keV}$ and a redshift
range of $z\simeq0.1-0.6$.  Thirty-eight of these systems come from
XXL-100-GC itself.  We combine the
\emph{XMM-Newton} survey data and the high-fidelity weak-shear catalog from
the CFHTLenS survey to obtain independent temperature and halo mass
measurements, respectively.  We describe the sample, data, and analysis,
 including details on the weak gravitational lensing analyses, in
Section~\ref{sec:Data}.  In Section~\ref{sec:MT} we present our main
results, the mass-temperature relation of XXL-100-GC.  We discuss a range of systematic uncertainties in our
analysis, confirming that they are sub-dominant to the statistical
uncertainties, in Section~\ref{sec:discussion}.  We also compare our
results with the literature in Section~\ref{sec:discussion}, and
summarise our results in Section~\ref{sec:Summary}.  We assume a WMAP9
\citep{Hinshaw2013} cosmology of $H_0=70\,{\rm
  km\,s^{-1}Mpc^{-1}}$, $\Omega_M=0.28$, and $\Omega_\Lambda=0.72$. All
statistical errors are reported to $68\%$ significance and upper
limits are stated at $3\sigma$ confidence.

\section{Sample, data and analysis}\label{sec:Data} 

\subsection{Survey and sample definition}

The XXL Survey is described in
detail by \citet[][Paper I, hereafter]{Pierre2016}.  This $\sim$50 deg$^2$
\emph{XMM-Newton} survey has a sensitivity of $\sim$
5$\times$10$^{-15}$ erg s$^{-1}$ cm$^{-2}$ in the [0.5-2] keV band
that provides a well-defined galaxy cluster sample for precision cosmology. The survey is an extension of the 11
deg$^2$ XMM-LSS survey \citep{Pierre2004} and consists of two 25
deg$^2$ areas.  The XXL-100-GC\footnote{XXL-100-GC data are available in computer readable form via the XXL Master Catalogue browser \url{http://cosmosdb.iasf-milano.inaf.it/XXL} and via the XMM XXL Database \url{http://xmm-lss.in2p3.fr}} sample is a flux-limited sample based on 100
clusters ranked brightest in flux. It is described in detail by \citet[][Paper II, hereafter]{Pacaud2016}, some of these clusters have previously been described in the XMM-LSS and XMM-BCS studies \citep{Clerc2014,
  Suhada2012}. We note that five systems (\texttt{XLSSC\,113, 114, 115, 550, and 551}) were observed in bad pointings that are contaminated by flaring. Subsequently, the sample was
supplemented with five additional clusters: \texttt{XLSSC\,091, 506, 516, 545 and
548}.  All systems within the XXL-100-GC sample are characterised as either
C1 or C2 \citep{Clerc2014}. The C1 objects have a high likelihood of detection 
and extension. The probability of contamination by spurious detection or
point sources for these systems is low ($<3\%$), whereas the C2 objects have
$\sim50\%$ contamination. The XXL-100-GC sample is estimated to be more
than $99\%$ complete down to  $\sim3\times$10$^{-14}$erg s$^{-1}$ cm$^{-2}$ and to have spectroscopic redshifts of $0.05\le
z\le1.07$ \citepalias{Pacaud2016}.

The mass-temperature relation presented in this paper is based on
weak-lensing mass measurements using the Canada-France-Hawaii
Telescope Lensing Survey (CFHTLenS) shear
catalogue\footnote{www.cfhtlens.org} \citep{Heymans2012,Erben2013}.
CFHTLenS spans a total survey area of $\sim154{\rm deg}^2$ that has
considerable overlap with the northern XXL field
(Fig.~\ref{fig:CFHTfield}).  Their shear catalogue comprises galaxy
shape measurements for a source density of 17 galaxies per
arcmin$^{2}$, as well as $u^*g' r' i' z'$-band photometry and photometric
redshifts for the same galaxies. The median photometric redshift of
the galaxies in the catalogue is $z_{\rm median}=0.75$
\citep{Hildebrandt2012}.

\begin{figure*}
  \centerline{
    \hspace*{-5mm}
    \includegraphics[width=95mm]{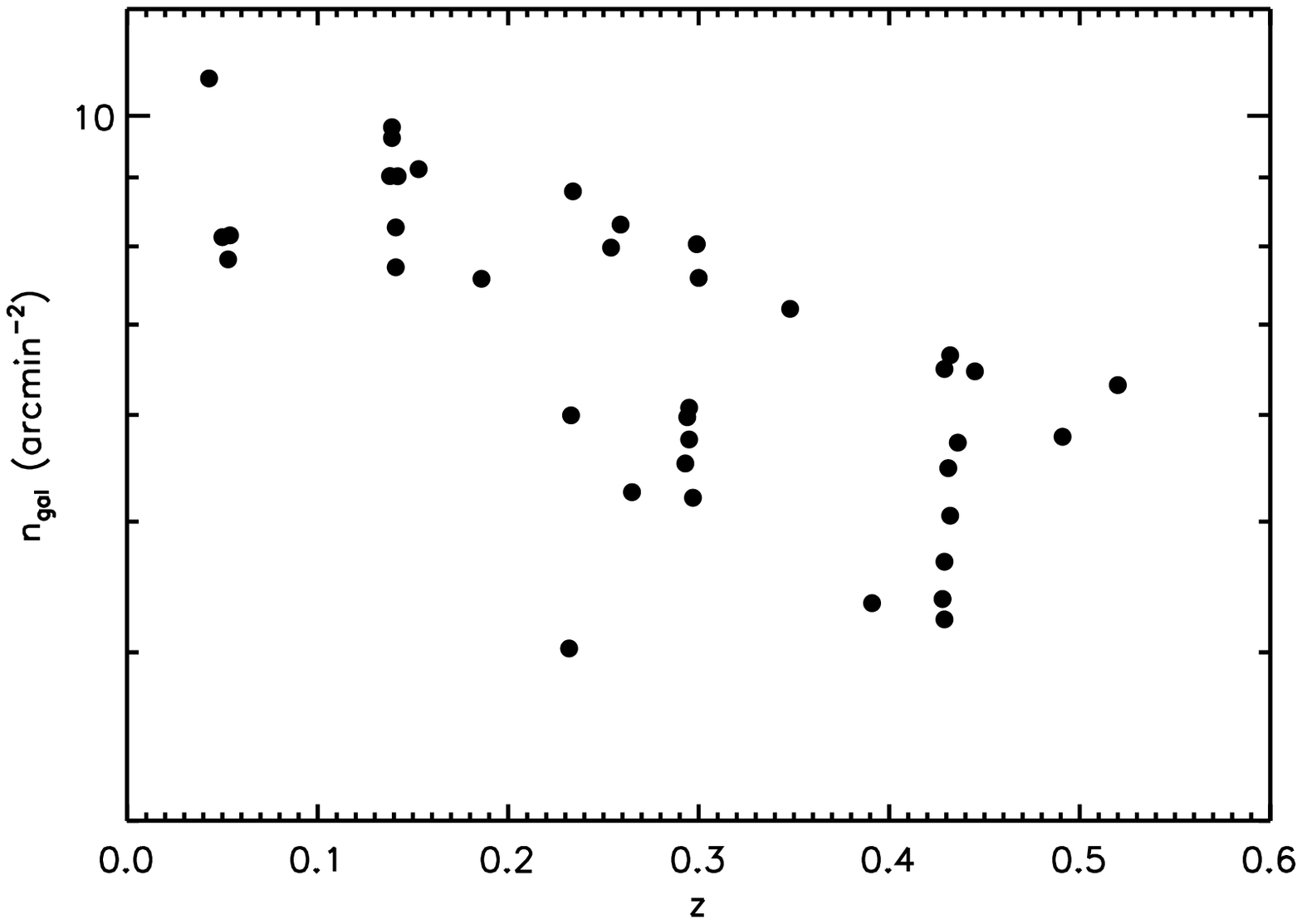}
    \hspace*{-5mm}
    \includegraphics[width=95mm]{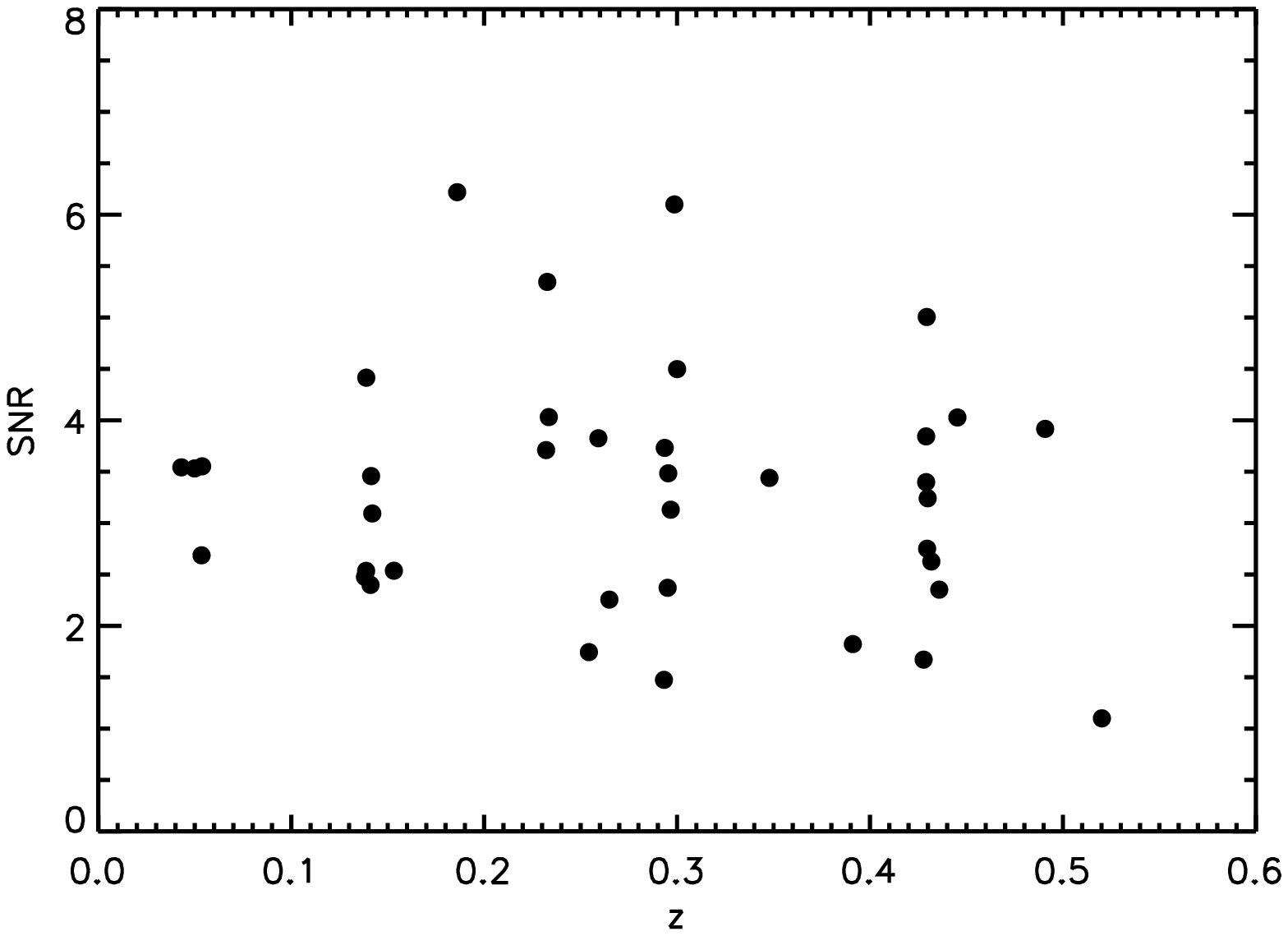}
  }
  \caption{{\sc Left}: Number density of background galaxies behind
    each galaxy cluster versus cluster redshift. 
       {\sc Right}: Weak-lensing shear
    signal-to-noise ratio as a function of cluster
    redshift. \label{fig:wlsnr}}
\end{figure*}

Fifty-two of the 100 XXL-100-GC sources lie in the northern XXL field, of
which 45 lie within the CFHTLenS survey area
(Fig.~\ref{fig:CFHTfield}).  A few of these 45 clusters lie at
redshifts beyond the median redshift of the CFHTLenS shear catalogue,
thus significantly reducing the number density of galaxies behind
these distant clusters.  We therefore limit our analysis to clusters
at $z<0.6$, which corresponds to imposing a lower limit on the
source density of $\sim 4\,{\rm arcmin}^{-2}$ (Fig.~\ref{fig:wlsnr}).
This gives a total sample of 38 galaxy clusters for which we have a
redshift, faint galaxy shape measurements, and an X-ray temperature
(Table~\ref{tabmass}). All 38 of these galaxy clusters are
classified as C1 with the exception of \texttt{XLSSC114}, which is a C2 class
system.

\subsection{X-ray temperatures}\label{sec:tx}

\begin{figure}
  \hspace*{-0.8cm}
  \includegraphics[width=\linewidth]{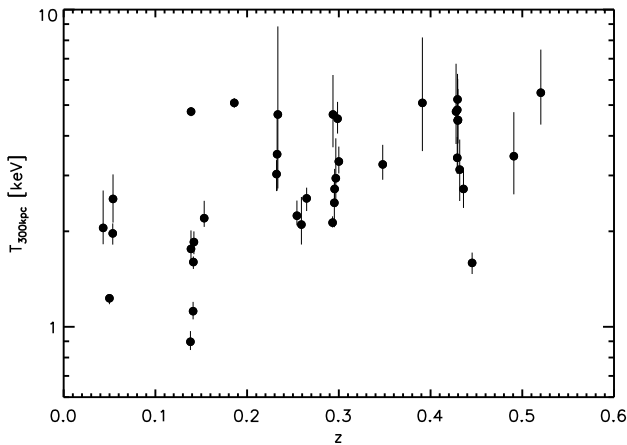}  \caption{Redshift versus X-ray temperature $T_{\rm 300kpc}$ for the 38 clusters
    from XXL-100-GC that are located within the
    CFHTLenS shear catalogue footprint. \label{fig:zTplane}}
\end{figure}

The temperature of the intracluster medium of each cluster is measured
and described in detail by \citet[][Paper III, hereafter]{Giles2016}.  Here
we summarise the key points pertaining to our analysis.

The spectra are extracted using a circular aperture of radius 0.3 Mpc centred on the X-ray positions, with a
minimum of 5 counts bin$^{-1}$. Point sources are identified using
SExtractor and excluded from the analysis; the images are visually inspected for any that might have been missed. Radial profiles of each source were extracted within the
$0.5-2\,{\rm keV}$ band with the background subtracted. The detection
radius was defined as the radius at which the source is detected to
$0.5\sigma$ above the background. Background regions were 
taken as annuli centred on the
observation centre with a width equal to the spectral extraction
region and the region within the detection radius excluded. Where this
was not possible, the background was measured from an annulus centred on
the cluster with inner radius set to the detection radius and outer
radius as 400~arcsec.

The X-ray temperatures span $1.1\,{\rm keV}\le  T_{\rm 300kpc}<5.5\,{\rm keV}$
(Figure \ref{fig:zTplane}) and are non-core excised owing to the limited angular
resolution of \emph{XMM-Newton}. The temperatures are extracted within
a fixed physical radius of $0.3\,{\rm Mpc}$ such that they are straightforward to
calculate from shallow survey data without needing to estimate
the size of the cluster.  This is the largest radius within
which it is possible to measure a temperature for the whole XXL-100-GC sample.  To
check the sensitivity of our main results to this choice of aperture,
we also re-fit the mass-temperature relation discussed in the results
section using the temperatures that are available in larger apertures
up to $0.5\,{\rm Mpc}$, and find that the systematic differences
between the respective fit parameters are negligible compared with the
statistical errors on the fits.

\subsection{Cool core strength}

The cool core strength of XXL-100-GC is estimated
by \citetalias{Democles2015} using the concentration
parameter method of \citet{Santos2008}.  We summarise a few key
points of the analysis here.  The X-ray surface brightness profile is
extracted within concentric annuli centred on the X-ray peak, it is
both background-subtracted and exposure corrected and then re-binned
to obtain a minimum signal-to-noise ratio (S/N) of 3 in each bin.  The
profiles are fit using three 3D density profile models which are
projected on the sky and convolved with the \emph{XMM-Newton} point
spread function (PSF).  Depending on the number of bins in the surface
brightness profile ($n_{\rm bin}$), a more or less flexible $\beta$-model
is fit to the data: $\beta=2/3$ is assumed for profiles with $n_{\rm
  bin}<3$; $\beta$ is a free parameter for $3\le n_{\rm bin}\le4$; a
double $\beta$ model is used for $n_{\rm bin}>4$.  The surface
brightness concentration parameter (CSB) is defined as the ratio of
the integrated profile within 40 kpc to that within 400 kpc,
CSB=SB($<$40 kpc)/SB($<$400 kpc).  The cool core status is defined as
\begin{itemize}
\item Non-cool core: CSB $<$ 0.075
\item Weak cool core: 0.075 $\le$ CSB $\le$ 0.155
\item Strong cool core: CSB $>$ 0.155
\end{itemize}

\subsection{Weak gravitational lensing}\label{sec:WL}

We use the full photometric redshift probability distribution, $P(z)$,
of each galaxy in the CFHTLenS shear catalogue to identify galaxies
behind our cluster sample.  Galaxies are selected as background
galaxies if they satisfy
\begin{equation}
	z_{s}-\delta{z_s}(3\sigma) > z + 0.01,
\end{equation}
where $z_s$ is the peak of the respective galaxy's $P(z)$, $z$ is
the cluster redshift, $\delta{z_s}(3\sigma)$ is the $99.7\%$ lower
confidence interval on $z_s$, and the last term represents a velocity
offset of $3000\,{\rm km\,s}^{-1}$ as a conservative allowance for the velocity width of
the cluster galaxy distributions.  

The method outlined in \cite{Velander2014} and \cite{Miller2013} is
used to calibrate the gravitational shear measurements.  The raw
ellipticity values ($e_1$, $e_2$) undergo two calibration
corrections, a mulitiplicative component ($m$) derived from
simulations \citep{Miller2013} and an additive component ($c$) derived
from the data \citep{Heymans2012}. The observed ellipticity can be
written as
\begin{equation}
	\rm e^{obs}=(1+m){\rm e}^{int}+c+\Delta {\rm e}
\end{equation}
where $\rm e^{int}$ is the intrinsic ellipticity and $\Delta
\rm e$ is the noise on the measurement.

The multiplicative component $m$ is dependent on both galaxy size and
S/N and gives, on average, a $6\%$ correction. The
additive component $c$ is similarly dependent on the galaxy size, and the
S/N determined by Lensfit. For the CFHTLenS data
$\langle c_1 \rangle$ is consistent with zero and $c_2$ is subtracted
from $e_2$ for each galaxy. The multiplicative correction is applied
as an average ensemble of each bin.

A weighting is also applied that corrects for the geometry of the
lens-source system in the form of the lensing kernel $\xi=D_{\rm
  LS}/D_{\rm S}$, where $D_{\rm LS}$ and $D_{\rm S}$ are the angular
diameter distances between the lens and the source, and between the
observer and the source, respectively. This is applied as a ratio
between that of the cluster-galaxy system and that of the reference
$\eta=\xi/\xi_{\rm ref}$.  The reference is taken as the mode
source redshift of the sum of all background galaxy weighted $P(z_s)$,
i.e.\ the mode of
\begin{equation}
n(z_s)=\sum\limits^{N_{\rm gal}}_{i=1}w_iP_i(z_s)
\end{equation}
where $w_i$ is the CFHTLenS inverse variance weight \citep[][equation 8]{Miller2013} applied to
calibrate for the likelihood of the measured ellipticity and intrinsic shape noise.  The
calibrated shear at a distance $r$ from the cluster centre therefore takes the form
\begin{eqnarray}
	\langle \gamma(r)^{cal} \rangle
        &=&\frac{\sum\limits_{i=1}^{N_{\rm gal}} w_i \eta_i \gamma_i^{int} 
          \sum\limits_{i=1}^{N_{\rm gal}} w_i
          \eta_i}{\sum\limits_{i=1}^{N_{\rm gal}} w_i \eta_i (1+m_i)\sum
          \limits_{i=1}^{N_{\rm gal}} w_i \eta_i^2}\nonumber. \\
\end{eqnarray}

In the weak-lensing limit the shear can be estimated as the average
complex ellipticity $\gamma \approx \langle {\rm e} \rangle$, where
${\rm e} \equiv e_1 +\textbf{i}e_2$. In terms of tangential and cross-component ellipticity,
\begin{eqnarray}
	{\rm e}_+ &=& -\Re  e^{-2\textbf{i} \phi} = -(e_2-c_2)
        \sin(2\phi) - e_1 \cos(2\phi) \\ {\rm e}_\times &=& - \Im
         e^{-2\textbf{i} \phi} = e_1 \sin(2\phi) - (e_2-c_2),
        \cos(2\phi)
\end{eqnarray}
where the tangential shear, $\rm e_+(r)$, is the signal that can be
modelled in terms of the total matter density profile of the lens. The 
cross shear $\rm e_\times(r)$ is orientated 45$^\circ$ with respect to
the tangential component and should be consistent with zero as a
check on systematic errors.

We extract the shear profile of each cluster within a $0.15-3\,{\rm
  Mpc}$ annulus.  The inner radial cut helps to ameliorate centring
uncertainties, and the outer radial cut is motivated by numerical
simulations \citep{Becker2011}.  The cluster centre is taken as the
X-ray centroid.  For reference, the mean offset between the X-ray
centroid and the brightest cluster galaxy (BCG) is $\langle \delta
r\rangle=64.7\, {\rm kpc}$.  Our results are unchanged if we centre the
shear profiles on the respective BCGs (see section
\ref{sec:systematics} for more details).

The shear is binned in eight radial bins equally spaced in log and with a
lower limit of 50 galaxies per radial bin.  If this threshold is not
met, the bin is combined with the next radial bin.  The errors on the shear in each radial bin are estimated
from $10^3$ bootstrap resamples with replacement and includes the large scale structure covariance \citep{Schneider1998}:
\begin{equation}
	C_{ij}^{LSS}=\int P_k(l)J_2(l\theta_i)J_2(l\theta_j)\frac{ldl}{2\pi},
\end{equation}
where $P_k(l)$ is the weak-lensing power spectrum as a function of angular multipole $l$ and $J_2(l\theta)$ is the second-order Bessel function of the first type at radial bins $\theta_i$ and $\theta_j$.

Shear S/N is calculated following
\cite{Okabe2010} as
\begin{equation}
	{\rm (S/N)}^2 =
        \sum^{N_{bin}}_{n=1}\frac{\langle {\rm e}_{+} (r_n)
          \rangle^2}{\sigma_{\rm e+}^2(r_n)}.
\end{equation} 
For our sample the weak-lensing S/N ranges from $\rm
1\le S/N \le7$.  However we include all objects in the
mass-temperature relation regardless of the S/N value to avoid
imposing a low-shear selection on top of the original X-ray selection.

\begin{table*}
\begin{center}
\caption{Cluster properties and mass estimates.
\label{tabmass}}
\begin{tabular}{lrrrrrrrrrr}
\hline
\\
\multicolumn{1}{c}{Name} &
\multicolumn{1}{c}{z} &
\multicolumn{1}{c}{$T_{\rm300kpc}$} &
\multicolumn{1}{c}{c$_{200}$} &
\multicolumn{1}{c}{$M_{\rm 200,WL}$} &
\multicolumn{1}{c}{$M_{\rm 500,WL}$} &
\multicolumn{1}{c}{$r_{\rm 500,WL}$} &
\multicolumn{1}{c}{$\delta r$} &
\multicolumn{1}{c}{$\delta$r/$r_{\rm 500,WL}$} &
\multicolumn{1}{c}{CSB} &
\multicolumn{1}{c}{SNR} \\
  \multicolumn{1}{c}{} &
  \multicolumn{1}{c}{} &
  \multicolumn{1}{c}{(keV)} &
  \multicolumn{1}{c}{} &
  \multicolumn{1}{c}{($10^{14}h^{-1}_{70} \rm M_\odot $)} &
  \multicolumn{1}{c}{($10^{14}h^{-1}_{70} \rm M_\odot$)} &
    \multicolumn{1}{c}{(Mpc)} &
    \multicolumn{1}{c}{($10^{-2}$Mpc)} &
      \multicolumn{1}{c}{($10^{-1}$)} &
  \multicolumn{1}{c}{($10^{-2}$)} &
    \multicolumn{1}{c}{} \\
      \multicolumn{1}{c}{(1)} &
  \multicolumn{1}{c}{(2)} &
  \multicolumn{1}{c}{(3)} &
  \multicolumn{1}{c}{(4)} &
  \multicolumn{1}{c}{(5)} &
  \multicolumn{1}{c}{(6)} &
    \multicolumn{1}{c}{(7)} &
    \multicolumn{1}{c}{(8)} &
      \multicolumn{1}{c}{(9)} &
  \multicolumn{1}{c}{(10)} &
    \multicolumn{1}{c}{(11)} \\
\hline
\texttt{XLSSC	006}	&	0.429	&	4.8	$^{+	0.5	}_{-	0.4	}$&	2.7	&	5.3	$^{+	6.0	}_{-	2.3	}$&	3.4	$^{+	3.7	}_{-	1.4	}$&	0.9	$^{+	0.3	}_{-	0.2	}$&	10.1	&	1.1	&	8.0	$\pm$	1.0	&	3.4	\\
\texttt{XLSSC	011}	&	0.054	&	2.5	$^{+	0.5	}_{-	0.4	}$&	3.4	&	1.6	$^{+	2.0	}_{-	1.1	}$&	1.1	$^{+	1.3	}_{-	0.7	}$&	0.7	$^{+	0.2	}_{-	0.2	}$&	0.4	&	0.1	&	12.7	$\pm$	0.9	&	3.6	\\
\texttt{XLSSC	022}	&	0.293	&	2.1	$^{+	0.1	}_{-	0.1	}$&	3.4	&	0.5	$^{+	0.9	}_{-	0.4	}$&	0.4	$^{+	0.5	}_{-	0.2	}$&	0.5	$^{+	0.2	}_{-	0.1	}$&	4.5	&	1.0	&	34.6	$\pm$	2.6	&	1.5	\\
\texttt{XLSSC	025}	&	0.265	&	2.5	$^{+	0.2	}_{-	0.2	}$&	3.1	&	1.7	$^{+	1.6	}_{-	1.3	}$&	1.1	$^{+	1.0	}_{-	0.8	}$&	0.7	$^{+	0.2	}_{-	0.2	}$&	0.0	&	0.0	&	27.9	$\pm$	2.7	&	2.3	\\
\texttt{XLSSC	027}	&	0.295	&	2.7	$^{+	0.4	}_{-	0.3	}$&	2.9	&	3.3	$^{+	3.9	}_{-	2.1	}$&	2.1	$^{+	2.4	}_{-	1.4	}$&	0.8	$^{+	0.2	}_{-	0.2	}$&	8.1	&	1.0	&	4.7	$\pm$	2.5	&	3.5	\\
\texttt{XLSSC	041}	&	0.142	&	1.9	$^{+	0.1	}_{-	0.2	}$&	3.4	&	1.0	$^{+	0.9	}_{-	0.7	}$&	0.7	$^{+	0.6	}_{-	0.5	}$&	0.6	$^{+	0.1	}_{-	0.2	}$&	1.3	&	0.2	&	29.9	$\pm$	2.5	&	3.1	\\
\texttt{XLSSC	054}	&	0.054	&	2.0	$^{+	0.2	}_{-	0.2	}$&	3.5	&	1.1	$^{+	1.6	}_{-	0.7	}$&	0.7	$^{+	1.1	}_{-	0.5	}$&	0.6	$^{+	0.2	}_{-	0.2	}$&	0.5	&	0.1	&	11.1	$\pm$	1.3	&	2.7	\\
\texttt{XLSSC	055}	&	0.232	&	3.0	$^{+	0.3	}_{-	0.3	}$&	2.8	&	8.1	$^{+	7.6	}_{-	3.1	}$&	5.2	$^{+	4.7	}_{-	2.0	}$&	1.1	$^{+	0.3	}_{-	0.2	}$&	4.2	&	0.4	&	11.3	$\pm$	1.9	&	3.7	\\
\texttt{XLSSC	056}	&	0.348	&	3.2	$^{+	0.5	}_{-	0.3	}$&	2.8	&	4.5	$^{+	2.7	}_{-	2.4	}$&	2.8	$^{+	1.7	}_{-	1.5	}$&	0.9	$^{+	0.2	}_{-	0.2	}$&	6.4	&	0.7	&	5.6	$\pm$	1.7	&	3.4	\\
\texttt{XLSSC	057}	&	0.153	&	2.2	$^{+	0.3	}_{-	0.1	}$&	3.7	&	$\le$ 0.9				    &	$\le$ 0.6				    &	$\le$ 0.6				    &	3.0	&	0.7	&	17.1	$\pm$	1.8	&	2.5	\\
\texttt{XLSSC	060}	&	0.139	&	4.8	$^{+	0.2	}_{-	0.2	}$&	3.2	&	2.1	$^{+	1.4	}_{-	1.5	}$&	1.4	$^{+	0.9	}_{-	1.0	}$&	0.8	$^{+	0.1	}_{-	0.3	}$&	13.5	&	1.8	&	2.3	$\pm$	0.1	&	4.4	\\
\texttt{XLSSC	061}	&	0.259	&	2.1	$^{+	0.5	}_{-	0.3	}$&	2.9	&	3.8	$^{+	0.9	}_{-	2.1	}$&	2.4	$^{+	0.5	}_{-	1.3	}$&	0.9	$^{+	0.1	}_{-	0.2	}$&	2.9	&	0.3	&	9.9	$\pm$	3.3	&	3.8	\\
\texttt{XLSSC	083}	&	0.430	&	4.5	$^{+	1.1	}_{-	0.7	}$&	2.7	&	4.0	$^{+	3.6	}_{-	2.8	}$&	2.5	$^{+	2.2	}_{-	1.7	}$&	0.8	$^{+	0.2	}_{-	0.3	}$&	4.1	&	0.5	&	7.0	$\pm$	2.4	&	3.2	\\
\texttt{XLSSC	084}	&	0.430	&	4.5	$^{+	1.6	}_{-	1.3	}$&	2.7	&	4.3	$^{+	3.2	}_{-	3.2	}$&	2.7	$^{+	1.9	}_{-	2.0	}$&	0.9	$^{+	0.2	}_{-	0.3	}$&	10.9	&	1.3	&	3.0	$\pm$	0.7	&	2.8	\\
\texttt{XLSSC	085}	&	0.428	&	4.8	$^{+	2.0	}_{-	1.0	}$&	3.2	&	$\le$ 2.6				    &	$\le$ 1.21				    &	$\le$ 0.7				    &	0.0	&	0.0	&	10.6	$\pm$	4.3	&	1.7	\\
\texttt{XLSSC	087}	&	0.141	&	1.6	$^{+	0.1	}_{-	0.1	}$&	3.6	&	0.5	$^{+	0.4	}_{-	0.4	}$&	0.3	$^{+	0.3	}_{-	0.2	}$&	0.5	$^{+	0.1	}_{-	0.2	}$&	0.9	&	0.2	&	41.5	$\pm$	2.9	&	3.5	\\
\texttt{XLSSC	088}	&	0.295	&	2.5	$^{+	0.6	}_{-	0.4	}$&	3.1	&	1.8	$^{+	1.3	}_{-	1.5	}$&	1.2	$^{+	0.9	}_{-	0.9	}$&	0.7	$^{+	0.1	}_{-	0.3	}$&	28.2	&	4.2	&	2.7	$\pm$	0.4	&	2.4	\\
\texttt{XLSSC	090}	&	0.141	&	1.1	$^{+	0.1	}_{-	0.1	}$&	4.1	&	$\le$ 0.6				    &	$\le$ 1.2				    &	$\le$ 0.7			    &	0.9	&	0.3	&	41.7	$\pm$	4.2	&	2.4	\\
\texttt{XLSSC	091}	&	0.186	&	5.1	$^{+	0.2	}_{-	0.2	}$&	2.8	&	9.7	$^{+	3.3	}_{-	2.9	}$&	6.2	$^{+	2.1	}_{-	1.8	}$&	1.2	$^{+	0.1	}_{-	0.1	}$&	5.0	&	0.4	&	2.5	$\pm$	0.1	&	6.2	\\
\texttt{XLSSC	092}	&	0.432	&	3.1	$^{+	0.8	}_{-	0.6	}$&	3.2	&	$\le$ 2.2				    &	$\le$ 1.4				    &	$\le$ 0.7				    &	26.3	&	7.9	&	6.9	$\pm$	1.7	&	2.6	\\
\texttt{XLSSC	093}	&	0.429	&	3.4	$^{+	0.6	}_{-	0.4	}$&	2.7	&	5.9	$^{+	3.5	}_{-	3.0	}$&	3.7	$^{+	2.1	}_{-	1.8	}$&	0.9	$^{+	0.2	}_{-	0.2	}$&	2.9	&	0.3	&	5.4	$\pm$	1.6	&	3.8	\\
\texttt{XLSSC	095}	&	0.138	&	0.9    $^{+  0.1   }_{-  0.1 }$&    3.6 	& $\le$	1.0				    & $\le$ 0.6 				    & $\le$0.6		    &	0.0	&	0.0			&  	40.3  $\pm$  14.9 & 2.5 \\
\texttt{XLSSC	096}	&	0.520	&	5.5	$^{+	2.0	}_{-	1.1	}$&	3.5	&	$\le$ 1.4				    &	$\le$0.9				    &	$\le$0.6				    &	5.0	&	1.7	&	7.3	$\pm$	2.5	&	1.1	\\
\texttt{XLSSC	098}	&	0.297	&	2.9	$^{+	1.0	}_{-	0.6	}$&	3.0	&	2.8	$^{+	3.6	}_{-	2.3	}$&	1.8	$^{+	2.3	}_{-	1.5	}$&	0.8	$^{+	0.2	}_{-	0.3	}$&	2.3	&	0.3	&	17.1	$\pm$	6.7	&	3.1	\\
\texttt{XLSSC	099}	&	0.391	&	5.1	$^{+	3.1	}_{-	1.5	}$&	3.5	&	$\le$ 2.2				    &	$\le$ 1.4				    &	$\le$ 0.7				    &	1.9	&	0.6	&	6.6	$\pm$	1.8	&	1.8	\\
\texttt{XLSSC	103}	&	0.233	&	3.5	$^{+	1.2	}_{-	0.8	}$&	2.8	&	8.5	$^{+	4.2	}_{-	3.0	}$&	5.4	$^{+	2.6	}_{-	1.8	}$&	1.1	$^{+	0.2	}_{-	0.2	}$&	4.2	&	0.4	&	6.9	$\pm$	2.6	&	5.3	\\
\texttt{XLSSC	104}	&	0.294	&	4.7	$^{+	1.5	}_{-	1.0	}$&	3.0	&	2.6	$^{+	4.1	}_{-	1.3	}$&	1.7	$^{+	2.6	}_{-	0.9	}$& 	0.8	$^{+	0.3	}_{-	0.2	}$&	14.9	&	2.0	&	9.9	$\pm$	3.7	&	3.7	\\
\texttt{XLSSC	105}	&	0.429	&	5.2	$^{+	1.1	}_{-	0.8	}$&	2.4	&	19.8$^{+	6.5	}_{-	7.7	}$&	12.1	$^{+	3.9	}_{-	4.6	}$&	1.4	$^{+	0.1	}_{-	0.2	}$&	14.3	&	1.0	&	3.5	$\pm$	0.7	&	5.0	\\
\texttt{XLSSC	106}	&	0.300	&	3.3	$^{+	0.4	}_{-	0.3	}$&	2.8	&	6.8	$^{+	3.0	}_{-	3.3	}$&	4.3	$^{+	1.8	}_{-	2.1	}$&	1.0	$^{+	0.1	}_{-	0.2	}$&	27.2	&	2.6	&	7.0	$\pm$	1.3	&	4.5	\\
\texttt{XLSSC	107}	&	0.436	&	2.7	$^{+	0.4	}_{-	0.3	}$&	2.8	&	2.8	$^{+	4.8	}_{-	2.2	}$&	1.8	$^{+	3.0	}_{-	1.4	}$&	0.7	$^{+	0.3	}_{-	0.3	}$&	0.0	&	0.0	&	13.0	$\pm$	2.6	&	2.4	\\
\texttt{XLSSC	108}	&	0.254	&	2.2	$^{+	0.3	}_{-	0.2	}$&	3.9	&	$\le$ 0.9				    &   $\le$ 0.6				    &	$\le$ 0.5				    &	4.0	&	1.3	&	14.0	$\pm$	2.5	&	1.7	\\
\texttt{XLSSC	109}	&	0.491	&	3.5	$^{+	1.3	}_{-	0.8	}$&	2.6	&	7.6	$^{+	6.6	}_{-	4.5	}$&	4.7	$^{+	4.0	}_{-	2.8	}$&	1.0	$^{+	0.2	}_{-	0.3	}$&	3.1	&	0.3	&	60.5	$\pm$	19.7	&	3.9	\\
\texttt{XLSSC	110}	&	0.445	&	1.6	$^{+	0.1	}_{-	0.1	}$&	2.7	&	4.6	$^{+	5.3	}_{-	1.6	}$&	2.9	$^{+	3.2	}_{-	1.0	}$&	0.9	$^{+	0.2	}_{-	0.1	}$&	17.7	&	2.0	&	2.6	$\pm$	0.4	&	4.0	\\
\texttt{XLSSC	111}	&	0.299	&	4.5	$^{+	0.6	}_{-	0.5	}$&	2.7	&	10.1	$^{+	3.0	}_{-	2.9	}$&	6.3	$^{+	1.8	}_{-	1.8	}$&	1.2	$^{+	0.1	}_{-	0.1	}$&	1.6	&	0.1	&	13.8	$\pm$	4.5	&	6.1	\\
\texttt{XLSSC	112}	&	0.139	&	1.8	$^{+	0.2	}_{-	0.2	}$&	3.4	&	1.2	$^{+	0.9	}_{-	0.8	}$&	0.8	$^{+	0.6	}_{-	0.5	}$&	0.6	$^{+	0.1	}_{-	0.2	}$&	6.9	&	1.1	&	9.3	$\pm$	1.5	&	2.5	\\
\texttt{XLSSC	113}	&	0.050	&	1.2	$^{+	0.0	}_{-	0.1	}$&	3.9	&	0.4	$^{+	0.6	}_{-	0.2	}$&	0.3	$^{+	0.4	}_{-	0.2	}$&	0.5	$^{+	0.2	}_{-	0.1	}$&	0.4	&	0.1	&	19.4	$\pm$	2.9	&	3.5	\\
\texttt{XLSSC	114}	&	0.234	&	4.7	$^{+	4.2	}_{-	1.9	}$&	3.1	&	2.1	$^{+	1.9	}_{-	1.0	}$&	1.4	$^{+	1.2	}_{-	0.6	}$&	0.7	$^{+	0.2	}_{-	0.1	}$&	5.5	&	0.8	&	5.0	$\pm$	1.9	&	4.0	\\
\texttt{XLSSC	115}	&	0.043	&	2.1	$^{+	0.6	}_{-	0.2	}$&	4.3	&	$\le$ 0.6				    &	$\le$ 0.4				    &	$\le$ 0.5				    &	2.5	&	0.8	&	6.9	$\pm$	2.3	&	3.5 \\	\hline
\end{tabular}
\bigskip

\begin{minipage}{0.9\linewidth}
{\footnotesize Column 1 is the cluster catalogue id number; Col. 2 is the
  cluster redshift; Col. 3  X-ray temperature measured within an
  aperture of 300 kpc; Col. 4 is the concentration parameter measured within
 $r_{\rm 200,WL}$; Cols. 5 and 6 are fitted estimates of weak-lensing mass centred
  on the X-ray centroid and measured within fitted $r_{\rm 200,WL}$ and $r_{\rm 500,WL}$
  respectively.  Upper limits on mass are given at 3 sigma
  confidence. Cols. 7 and 8 are the weak-lensing $r_{\rm 500,WL}$
  and the offset between the X-ray centroid
  and the BCG; Col. 9 is the  the BCG offset as a fraction of $r_{\rm 500,WL}$; Col. 10 is the CSB parameter and Col. 11 is the
  signal-to-noise ratio on the weak-lensing shear. Positions of the cluster X-ray centroids are listed in \citetalias{Pacaud2016} Table 1.}
\end{minipage}

\end{center}
\end{table*}

We model the shear profile as a \cite[][NFW hereafter]{Navarro1997} profile following the formalism set out by
\cite{Wright2000}.  A Markov chain Monte Carlo (MCMC) sampler with a
Gaussian likelihood is used to fit the NFW model to the shear
profile. The algorithm returns 5$\times10^{4}$ samples of the target
distribution using a jump proposal based on a Metropolis-Hastings
algorithm with a mean acceptance rate of 0.57.
The autocorrelation length is computed to thin correlated samples
within the chain and incorporates burn-in of 150 samples. The
Gelman-Rubin criterion \citep{Gelmanrubin1992} is computed for three chains
to ensure convergence.  The mass of each cluster is taken as the mode
of the posterior and the errors are given as $68\%$ credible regions
of the highest posterior density as this is the best representation of the skewed Gaussian posteriors.

Given the wide range of possible cluster mass, a uniform in log (Jeffreys) prior is
used to ensure scale invariance
$P(M|I)=\frac{1}{M\ln(10^{16}/10^{13})}$ ($10^{13} \le M_{\rm 200}
\le10^{16}M_\odot$).  Given the generally low-shear S/N, we fix
cluster concentration to values from a mass-concentration relation
based on N-body simulations \citep{Duffy2008}:
\begin{equation}
  c_{200}=5.71(1+z)^{-0.47}\left(
  \frac{M_{\rm 200}}{2\times10^{12}h^{-1}M_\odot}\right)^{-0.084}.
\end{equation}
We test the sensitivity of our results to the choice of this relation and find that it is not a dominant source of uncertainty (see section
4.1 for more details).

To estimate $M_{\rm \Delta,WL}$ for each cluster we integrate the NFW model out
to the radius at which the mean density of the halo is $\Delta \rho_{\rm
  crit}(z)$, where $z$ is the cluster reshift (Table
\ref{tabmass}) and $\Delta$=500:
\begin{eqnarray}
 {M}_{\rm \Delta, WL} &=& \int^{r_{\rm \Delta, WL}}_{0} \rho(r) 4\pi r^2 dr \nonumber \\ &=&
  4\pi\rho_sr_s^3\left[\ln\left(1+\frac{r_{\rm \Delta, WL}}{r_s}\right)-\frac{r_{\rm \Delta,WL}}{r_s+r_{\rm \Delta,WL}}\right]\nonumber .
  \\
\end{eqnarray}

\begin{figure*}
\includegraphics[width=\linewidth]{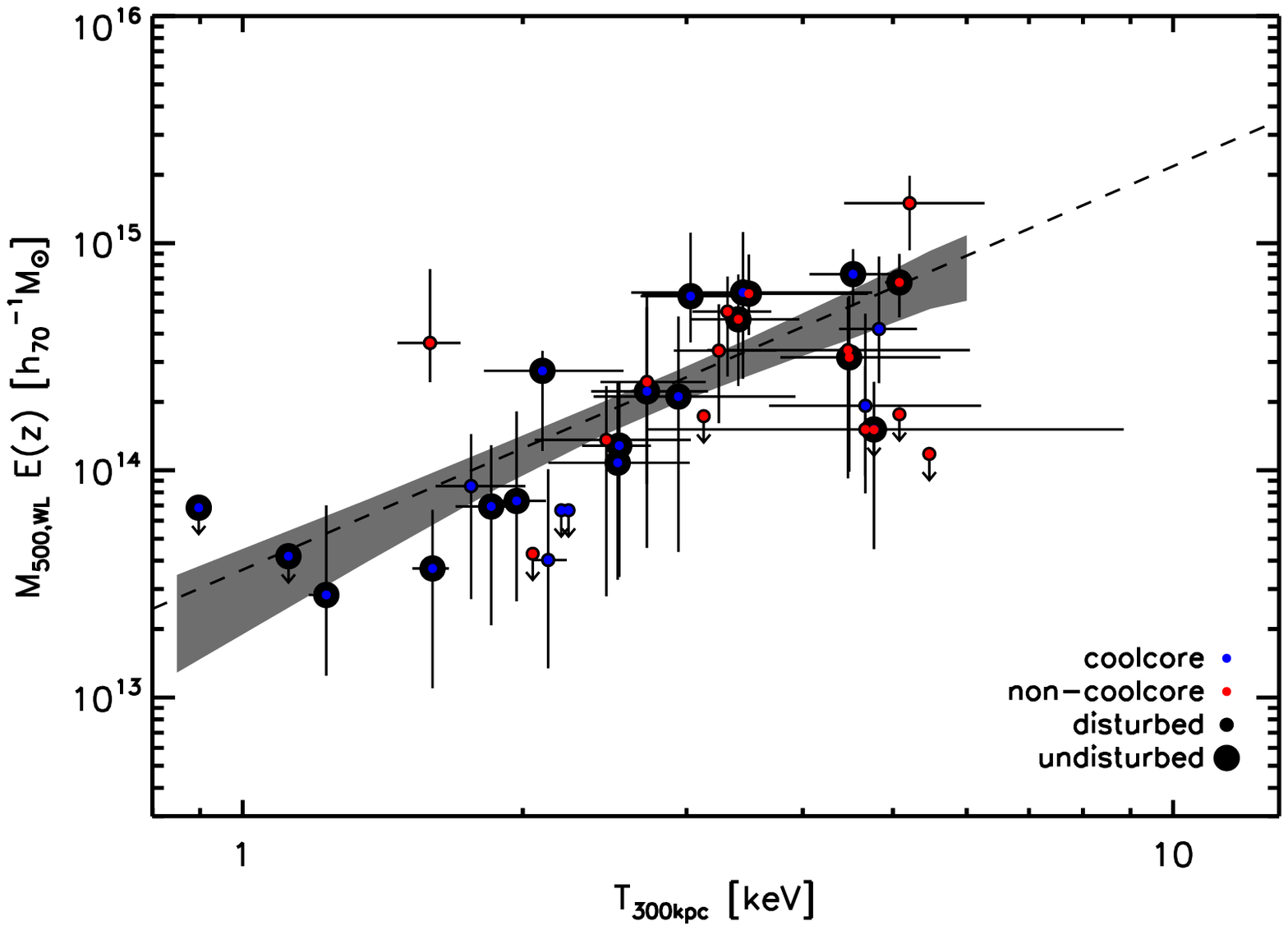}
\caption{The mass-temperature relation for 38 clusters drawn from
  XXL-100-GC for which weak-shear information is
  available from CFHTLenS. The line is the
  highest posterior density fit and the shaded region is the credible
  region.  Systems with upper limits on mass are indicated by arrows and plotted at 3 $\sigma$ confidence.
  \label{fig:MTrelationxxl}}
\end{figure*}

\section{Results}\label{sec:MT}

A positive correlation between our weak-lensing mass and X-ray
temperature measurements is evident (Figure \ref{fig:MTrelationxxl}).  In this section, we define
the scaling relation model that we will fit to the data, describe the
regression analysis, and present the main results.  We defer
consideration of possible systematic uncertainties and comparison with
the literature to section \ref{sec:discussion}.

\subsection{XXL mass-temperature relation}

\begin{figure*}
\includegraphics[width=\linewidth]{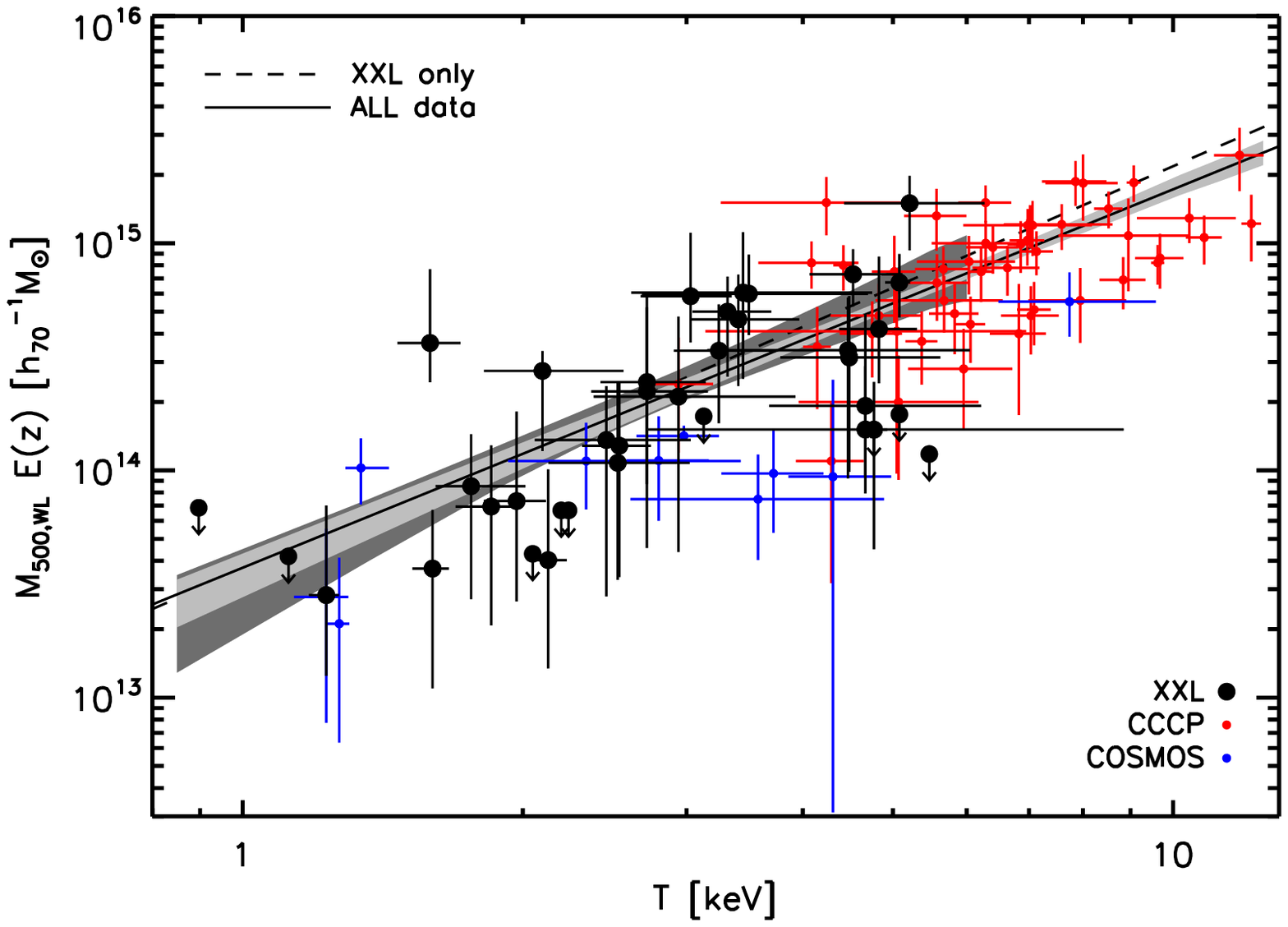}
\caption{Mass-temperature relation for the extended sample, including
  38 systems from XXL (black), 10 from COSMOS (blue), and 48 from CCCP
  (red). The solid line and light gray shaded region are the best fit
  scaling relation and $68\%$ credible interval for the
  XXL+COSMOS+CCCP sample. The dashed line and dark grey shaded region
  are the best fit and credible region for the XXL only
  sample. Systems with upper limits on mass are indicated by arrows and plotted at 3 sigma confidence.
  \label{fig:MTrelation}}
\end{figure*}

We model the mass--temperature relation as a power law:
\begin{equation}\label{MTrelationform}
 \log_{10} \left(\frac{M_{\rm 500}E(z)}{\rm M_\odot h_{70}^{-1}}\right)=a+b \log_{10} \left(\frac{T}{\rm keV}\right)
\end{equation}
with intercept $a$ and slope $b$, where
$E(z)=\sqrt{\Omega_m(1+z)^3+\Omega_\Lambda}$ describes the evolution
of the Hubble parameter.  We note that by not allowing any freedom in the
exponent of $E(z)$, we are assuming self-similar evolution.  This is
motivated by the large scatter which is apparent in our data, that
precludes us from constraining evolution at this time.

For the linear regression we use the Gibbs sampler implemented in
the multivariate Gaussian mixture model routine $\texttt{linmix\_err}$ \citep{Kelly2007} with the default of three Gaussians. 
We use 10$^{5}$ random draws of the sampler and take the fitted
parameters as the posterior mode and the error as the 68$\%$ highest
posterior density credible interval. When the number of data points is small, the Gibbs sampler will have difficulty  in reaching convergence. $\texttt{linmix\_err}$ also has the option of running as a Metropolis-Hastings algorithm, which is more efficient for small sample size. Tests implementing the Metropolis-Hastings algorithm give consistent results. 

We fit the model to the measured values of $M_{\rm 500,WL}$ and
$T_{\rm 300kpc}$.  For some galaxy clusters, the weak-lensing S/N
is so low that the we are only able to obtain an upper limit on
$M_{\rm 500,WL}$.  The posteriors of these systems are truncated by the lower bound prior on mass. 
Despite this, it is important to include these systems in the fit
because they are X-ray detected at high significance, and to exclude
them would add a further selection in addition to the primary X-ray
selection.  The fitting method used is able to incorporate upper limits as censored data using a likelihood that integrates over the censored and uncensored data separately \citep[see][for more details]{Kelly2007}. However their implementation is not suitable for our problem since we have prior knowledge of the X-ray detection we know that these systems should have a mass greater than 10$^{13}\rm M_\odot$, flagging them as censored data would contradict the mass prior used in fitting the NFW profile. Tests to recover scaling relation parameters on simulated toy data show that censoring leads to a positive bias in the slope. For systems where the lower credible region is truncated by the mass prior and hence underestimated we set the lower mass error equal to the upper mass error. In our toy model tests this gave the least bias in scaling relation parameters, with biases $\rm < 10\%$.

The mass-temperature relation based on the 38 clusters that overlap
between the XXL-100-GC and the CFHTLenS shear catalog has a slope of
$ b=1.78^{+0.37}_{-0.32}$, with an intrinsic scatter in natural log
of mass at fixed temperature of $\sigma_{\rm int \ln M|T}\simeq0.5$
(Table~\ref{tabparam}).

\begin{table*}
\begin{center}
\caption{Mass-temperature relation fit parameters for equation \ref{MTrelationform}. Fixed slope relations are denoted by FS.
\label{tabparam}}
{\renewcommand{\arraystretch}{1.3}
\begin{tabular}{lcccc}
\hline
\multicolumn{1}{c}{sample} &
\multicolumn{1}{c}{intercept} &
\multicolumn{1}{c}{slope} &
\multicolumn{1}{c}{intrinsic scatter} &
\multicolumn{1}{c}{N}\\
\multicolumn{1}{c}{} &
\multicolumn{1}{c}{($a$)} &
\multicolumn{1}{c}{($b$)} &
\multicolumn{1}{c}{($\sigma_{\mathrm{int \ln M|T}}$)} &
\multicolumn{1}{c}{}\\
\hline
XXL								&	13.56$^{+0.16}_{-0.17}$		&	1.78$^{+0.37}_{-0.32}$       	& 	0.53$^{+0.21}_{-0.17}$		&	38\\
XXL+COSMOS+CCCP 				&  	13.57$^{+0.09}_{-0.09}$ 		&	1.67$^{+0.14}_{-0.10}$       	&   	0.41$^{+0.07}_{-0.06}$ 		&	96\\
XXL FS							&	13.67$^{+0.07}_{-0.03}$		&	1.50					      	& 	0.48$^{+0.19}_{-0.08}$		&	38\\
XXL cool core						&	13.46$^{+0.19}_{-0.24}$		&	1.81$^{+0.43}_{-0.57}$	         & 	0.64$^{+0.26}_{-0.23}$		&	21\\
XXL non-cool core					&	14.18$^{+0.46}_{-0.39}$		&	0.75$^{+0.76}_{-0.73}$       	& 	0.50$^{+0.30}_{-0.22}$		&	17\\
XXL undisturbed					&	13.56$^{+0.15}_{-0.19}$		&	1.86$^{+0.35}_{-0.36}$       	& 	0.34$^{+0.25}_{-0.20}$		&	19\\
XXL disturbed						&	13.67$^{+0.40}_{-0.49}$		&	1.49$^{+0.82}_{-0.89}$       	& 	0.91$^{+0.28}_{-0.32}$		&	19\\
\hline
XXL cool core FS					&	13.59$^{+0.04}_{-0.08}$		&	1.50	         & 	0.72$^{+0.03}_{-0.16}$		&	21\\
XXL non-cool core FS				&	13.83$^{+0.04}_{-0.17}$		&	1.50 		& 	0.50$^{+0.15}_{-0.14}$		&	17\\
XXL undisturbed FS					&	13.71$^{+0.09}_{-0.08}$		&	1.50		&	0.39$^{+0.16}_{-0.13}$		&	19\\
XXL disturbed FS					&	13.62$^{+0.05}_{-0.12}$		&	1.50       	& 	0.75$^{+0.31}_{-0.16}$		&	19\\
\hline
\end{tabular}}

\bigskip

\begin{minipage}{0.9\linewidth}
{\footnotesize  }
\end{minipage}

\end{center}
\end{table*}

\subsection{Cool core status and dynamical disturbance}

We investigate whether the mass-temperature relation fit parameters
depend on the strength of cooling in the clusters cores and the
dynamical state of the clusters.  

First, we collectively classify weak and strong cool cores as cool
core systems and fit the mass-temperature relation to this cool core
subsample, and the non-cool core subsample.  The results of the fits
have large statistical uncertainties and intrinsic scatter.
The same is
true if we repeat the fits to the two subsamples holding the slope of
the respective relations fixed at the self-similar value of
$b=1.5$ (Table~2).

Second, we use the offset between the X-ray centroid and the BCG
\citep{Lavoie2015}, expressed as a fraction of $r_{\rm 500,WL}$, to
classify clusters as undisturbed $\delta r/r_{\rm 500,WL}<0.05$, and
disturbed $\delta r/r_{\rm 500,WL}>0.05$.  The scatter in the
mass-temperature relation for undisturbed clusters is less than that
of the disturbed clusters, albeit with large uncertainties. We see
similar results if we hold the slope of the relation fixed at
self-similar, as above.  This suggests that the disturbed clusters dominate the
scatter in the XXL-100 mass-temperature relation.  

It is tempting to attribute the large scatter in the mass-temperature
relation for disturbed clusters to the physics of the cluster merger
activity implied by a large value of $\delta r/r_{\rm 500,WL}$.  However we
caution that dynamically active clusters likely have more complicated
mass distributions than less active (``undisturbed'') clusters.  Our
ability to constrain reliable cluster mass measurements in the
$10^{13}<M_{500}<10^{14}M_\odot$ regime with low SNR survey data is
likely a function of the complexity of the mass distribution.  This mass range has not yet been explored to any great extent by
simulation studies (e.g.\ Becker \& Kravtsov 2011; Bah\'e et
al.\ 2012).  We will return to this question in a future article.

\subsection{Combination with other samples}

 \begin{figure}
\includegraphics[width=\linewidth]{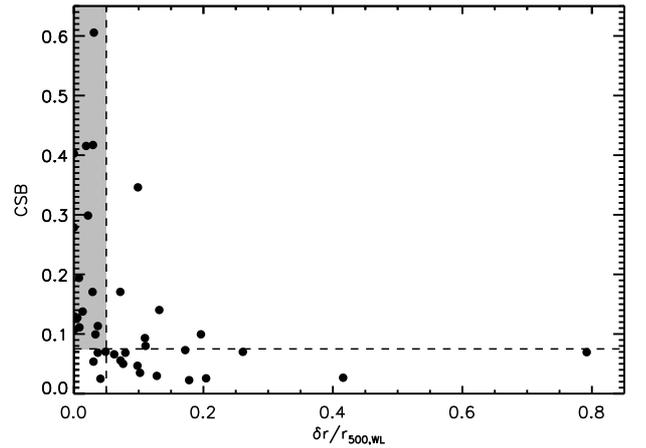}
\caption{CSB parameter versus the offset between X-ray centroid and
  BCG as a fraction of weak-lensing $r_{\rm 500,WL}$. The horizontal dashed
  line at CSB = 0.075 indicates the separation of cool core and
  non-cool core classed systems. The vertical dashed line at $\delta
  r/r_{\rm 500,WL}=0.05$ separates undisturbed and disturbed clusters. The
  grey shaded region shows the overlap between cool core and
  undisturbed clusters.
  \label{fig:csb_deltar}}
\end{figure}

\begin{figure}
\hspace*{-0.8cm}	
\includegraphics[width=\linewidth]{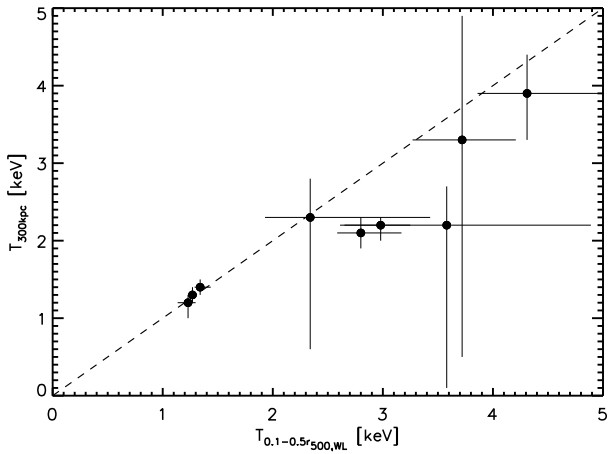}
\caption{Comparison of core excised X-ray temperatures
  \protect{\citep{Kettula2013}} and the re-derived temperatures measured within
  a 0.3Mpc aperture. The dashed line is
  equality. \label{fig:Tkettula_Tgiles}}
\end{figure}

To improve the precision and to extend the dynamic range of our
mass-temperature relation we now include 10 groups from COSMOS
\citep{Kettula2013} and 48 massive clusters from the Canadian Cluster
Comparison Project (CCCP; \cite{Mahdavi2013, Hoekstra2015}).  The COSMOS
groups are X-ray selected and their weak-lensing masses are based on deep
\emph{Hubble Space Telescope} observations, and follow a similar
analysis method to our own.  Unlike our sample, the temperatures of
the COSMOS systems are core excised.  We have therefore measured
non-core excised temperatures for the ten COSMOS groups within the same
$0.3{\rm Mpc}$ measurement aperture using the same analysis process described in section
\ref{sec:tx}.  Comparison between these non-core excised temperature
and the core excised temperatures used by \citet{Kettula2013} reveals
a bias of $\langle T_{\rm 300kpc}/T_{0.1-0.5r_{\rm 500,WL}}\rangle=0.91\pm0.05$ (Figure
\ref{fig:Tkettula_Tgiles}), and emphasise the importance of ensuring
that the temperatures are measured in a consistent manner when
combining samples.

We also obtained non-core excised temperatures for the CCCP clusters
analysed by \citet{Mahdavi2013} from the CCCP
web-site\footnote{http://sfstar.sfsu.edu/cccp/}, albeit within a
$0.5\,{\rm Mpc}$ aperture.  This is larger than the aperture that we
use for our own temperature measurements.  Given that the CCCP systems
are more massive than ours, we do not expect this difference in aperture
to have a significant affect on our results.  We confirm that this is
indeed the case (see section 4.1 for more details).

We fit the mass-temperature relation to the joint data set following
the same procedure as applied to the XXL-only sample in \S3.1.  The
statistical precision of the fit is much higher than that of the
XXL-only fit, and has very similar central values for all fit parameters
between the two fits (Table~3). The slope parameter of the joint fit
is $b=1.67^{+0.14}_{-0.10}$ with an intrinsic scatter of
$\sigma_{{\rm int}(\ln\,M\,|\,T)}=0.41^{+0.07}_{-0.06}$. 

\subsection{Mass estimates for XXL-100-GC}

The mass of each member of XXL-100-GC is computed
from the joint XXL+COSMOS+CCCP mass-temperature relation (see Table
\ref{tabparam}).  The uncertainties on these masses are estimated by
propagating uncertainties on individual temperature measurements, and the
 intrinsic scatter on the mass-temperature relation. The masses are presented in
\citetalias{Pacaud2016}, and denoted as $M_{\rm 500,MT}$ to indicate that they are based on the mass--temperature scaling relation.

\section{Discussion}\label{sec:discussion}

In \S\ref{sec:systematics} we discuss the effect of systematic
uncertainties on our results, and in \S\ref{sec:literature} we compare
our results with the literature.

\begin{figure*}
  \centerline{
    \hspace{-5mm}
    \includegraphics[width=95mm]{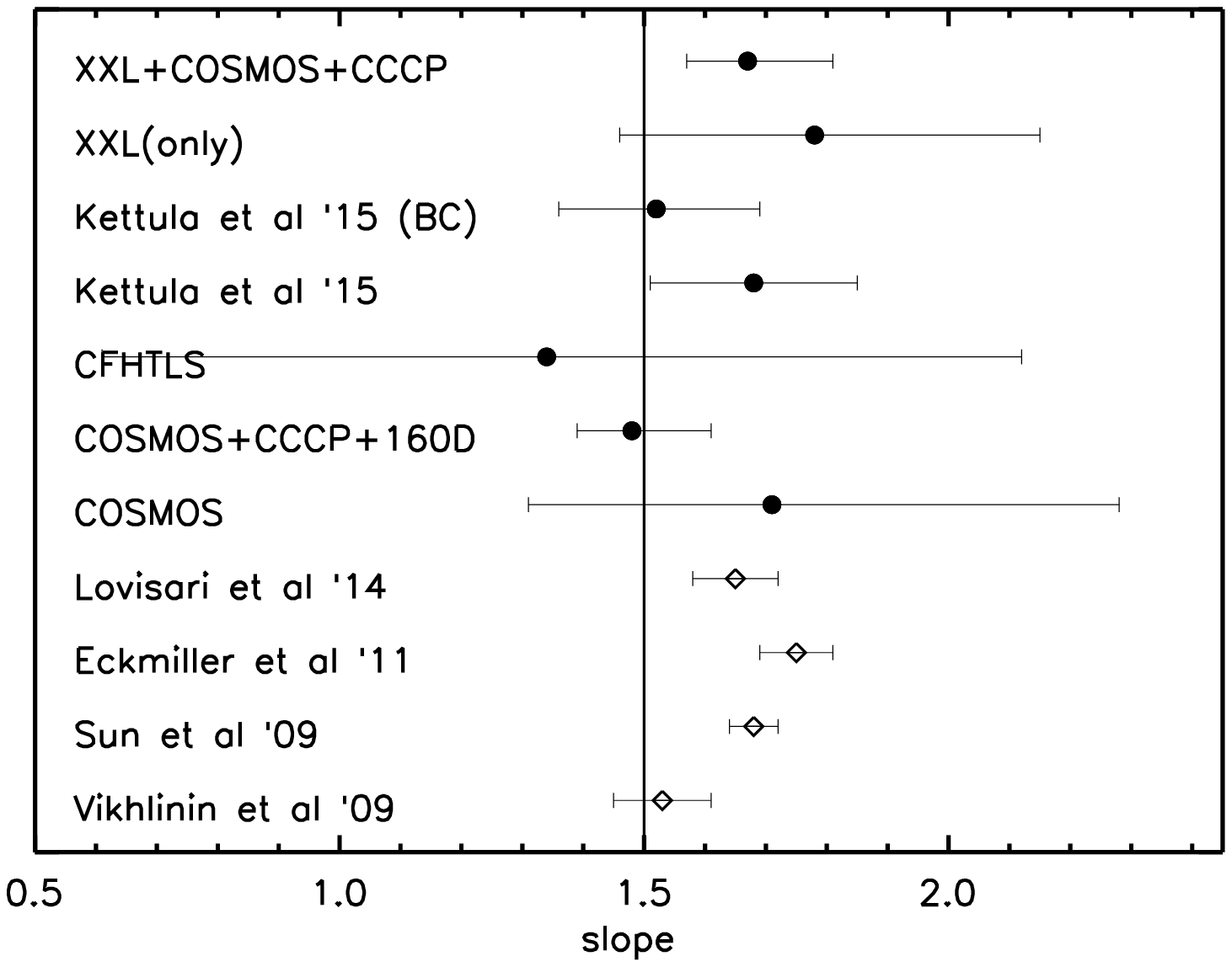}
    \hspace{-5mm}
    \includegraphics[width=95mm]{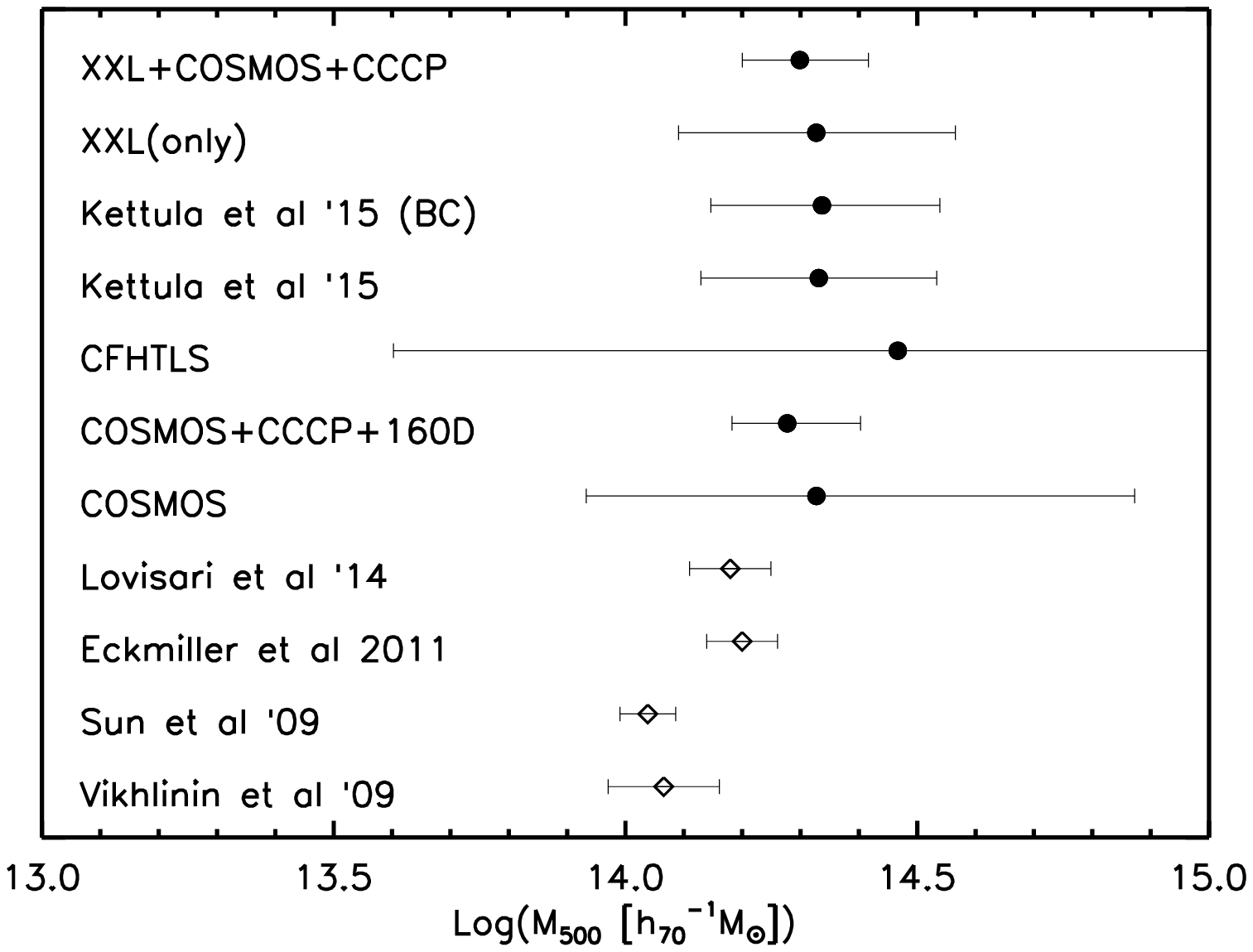}
  }
  \caption{{\sc Left}: Comparison of our results on the slope of the
    mass-temperature relation with those in the literature \citep{Eckmiller2011,Lovisari2015, Sun2009, Vikhlinin2009}. {\sc
      Right}: Comparison of the mass of a cluster of temperature
    $T=3\,{\rm keV}$ at $z=0.3$ based on mass-temperature relations and
    those in the literature.  In both panels, filled circles are
    samples that use weak-lensing masses, open diamonds are samples
    that use hydrostatic masses. The COSMOS+CCCP+160D and COSMOS-only relations are from \protect{\cite{Kettula2013} and the CFHTLS relation from \protect{\cite{Kettula2015}}. BC has been corrected for Eddington bias. \label{fig:compare}}}
\end{figure*}

\subsection{Systematic uncertainties}\label{sec:systematics}

Several sources of systematic uncertainty have been discussed in
the preceding sections.  Here we describe the tests that were performed to
assess the amplitude of these uncertainties.  

\medskip

\noindent\emph{Fitting method} -- We tested the robustness of the
fitting method on the resultant scaling parameters using {\sc
  mpfitexy} \citep{Williams2010}. This is a variation of the standard
\textsc{idl} fitting technique {\sc mpfit} \citep{Markwardt2009} that
minimises a $\chi^2$ statistic and iteratively adjusts for intrinsic
scatter. However, it does not calculate the error on the intrinsic scatter.
Using {\sc mpfitexy} the XXL+COSMOS+CCCP fit of 96 objects produces a
slope of $b=1.71\pm0.11$, intercept of $a = 13.55\pm0.09$,
and intrinsic scatter of $\sigma_{{\rm int}\ln M|T} = 0.38$, i.e.\ fully
consistent with our results presented in section \ref{sec:MT} (Table
\ref{tabparam}).

\medskip

\noindent\emph{Upper limits} -- To test the sensitivity of our results
to the treatment of clusters with upper limits on $M_{\rm 500,WL}$ we re-fitted
the mass-temperature relation excluding these objects, obtaining a
marginally shallower slope of $b=1.63\pm0.13$ and an intrinsic
scatter of $\sigma_{\ln M|T}=0.39\pm0.06$ for the joint
XXL+CCCP+COSMOS sample and $b=1.84\pm0.38$, $\sigma_{\ln M|T}=0.30\pm0.18$ for the XXL-only sample -- again, consistent with our main results. 

\medskip

\noindent\emph{Centring of the shear profile} -- Cluster masses are
dominated by statistical noise such that whether we centre the shear
profile on the BCG or on the X-ray centroid does not lead to a
large systematic uncertainty.  There is large scatter between the
masses derived from the different centres; however, the bias is minimal
($\langle M_{\rm 500,WL}^{\rm Xray}/M_{\rm 500,WL}^{\rm BCG}\rangle=1.00\pm0.16$) and so
does not have an impact on our results.  The BCG centred fits return a XXL-CCCP-COSMOS combined MT relation with slope $b=1.61\pm0.14$ and an intrinsic scatter of $\sigma_{{\rm int}\ln M|T} = 0.43 \pm 0.06$.

\medskip

\noindent\emph{Source selection} -- The photometric redshift uncertainty of galaxies and its contribution to the mass estimation of clusters in our sample is small $\langle d \xi/\xi \rangle = 0.13$ and so we used all background galaxies
with $P(z)$ measurements that satisfy our redshift cuts (Section
\ref{sec:WL}). \cite{Benjamin2013} use tests with spectroscopic
redshifts to find that within the CFHTLenS catalogue the redshifts are
most reliable between $0.1<z<1.3$. This is due to a fundamental
degeneracy in the angular cross-correlation method. At $z<0.1$, their
contamination model tends to underpredict contamination by
higher redshift galaxies. At $z>1.3$ the predicted contamination by
lower redshift galaxies is also underestimated.  We compared masses
derived using all galaxies to masses restricted to the reliable
redshift range $0.1<z<1.3$. The masses are impervious to the two source
selections with a ratio of $\langle M_{\rm 500,WL}^{0.1<z<1.3}/M_{\rm 500,WL}\rangle=1.13\pm0.18$. In our sample only 10$\%$ of the systems include the $z<0.1$ contaminated galaxies and the low number of $z>1.3$ galaxies should contribute little to the shear. This in combination with the large statistical uncertainties on shear would explain the agreement.

\medskip

\noindent\emph{Outer fitting radius} -- The systems considered in this
article are lower mass than most of those considered by \citet{Becker2011}.  Thus the outer radius to which the NFW model is
fitted to the measured shear profile may extend further into the
infall region than in their simulation study, and thus might bias our
mass measurements.  We implemented a simple test whereby we compared
the mass obtained from NFW models fitted to the annulus $0.15-2\,{\rm
  Mpc}$ to those described in section 2.4.  The mean ratio of the
masses derived from these fits and those upon which our results are
based (0.15 -- 3 Mpc) is $1.01\pm0.17$.

\medskip

\noindent\emph{Choice of mass-concentration relation} -- We adopted
the \cite{Duffy2008} mass-concentration relation for our mass modelling
of the shear signal, which aids comparison with the literature
\citep{Kettula2013}.  However observational studies
\citep[e.g.][]{Okabe2013,Umetsu2014} indicate that clusters are more
concentrated than expected from simulations  \citep[e.g.][]{Duffy2008, Bhattacharya2013}. \cite{Hoekstra2012} show that a 20$\%$
change in normalisation of the mass-concentration relation would bias
NFW-based masses by $\sim5-15\%$, although recent work by
\cite{Sereno2015} suggest the bias could be accounted for by selection
effects.  As a simple test, we perturbed the normalisation of the
\cite{Duffy2008} relation by a factor of 1.31 to bring it into line
with the stacked weak-lensing analysis of \cite{Okabe2013}.  The
masses that we computed using this perturbed relation are slightly
lower than our Duffy-based masses, although consistent within the
errors: $\langle M_{\rm Perturbed}/M_{\rm Duffy}\rangle=0.93\pm0.14$.
Although it is possible to obtain a mass when allowing concentration to be a free
parameter ($\langle M_{\rm free}/M_{\rm Duffy}\rangle=0.87\pm0.14$), we did not do this as we were not able to constrain concentration with this data. 
 The slope of the mass-temperature relation fits to the joint sample,
based on our perturbed and free-concentration masses are $b_{\rm
  perturbed}=1.75 \pm 0.13$ and $b_{\rm free}=1.71 \pm 0.14$. Within the errors both
are consistent with the Duffy concentration prior results. The XXL-only M--T relation using free-concentration masses
has regression parameters  $b=1.77\pm0.37$, $a=13.54\pm0.21$, and $\sigma_{\ln M|T}=0.38\pm0.20$.

\medskip

\noindent\emph{Cosmic shear test} -- \cite{Heymans2012} compute the
star-galaxy cross-correlation function of objects within the CFHTLenS
catalogue finding an amplitude much higher than expected from
simulations. Approximately 25$\%$ the fields fail this cosmic
shear test and when rejected bring the observations back into
agreement with simulations. This affects $\sim40\%$ of our systems: \texttt{XLSSC\,054, 055, 060, 056, 091, 095, 096, 098, 099, 103, 104, 105, 107,
108, 110, and 111}.  Excluding these systems from our sample does not
significantly change our results; for example a joint fit to the
remaining XXL clusters, COSMOS, and CCCP (80 systems in total)
yields $a=13.43^{+0.13}_{-0.09}$, $b=1.79^{+0.16}_{-0.12}$,
$\sigma_{int, \ln M|T}=0.42^{+0.07}_{-0.06}$. This suggests that it has an insignificant effect on cluster lensing where PSF residuals are reduced from the radial averaging. All CFHTLenS fields are used in both \cite{Velander2014} and \cite{Kettula2015}.  

\medskip

\noindent\emph{Mismatch in temperature measurement apertures} -- As
discussed in the results section, our temperature measurement aperture
differs from that used by CCCP.  This should not dramatically affect
our results as the temperature profile of clusters is shallow and for
groups 0.3 Mpc is a significant fraction of $r_{\rm 500,WL}$, whereas for the
massive clusters in CCCP the same holds at 0.5Mpc. Nonetheless, as a
test we computed temperatures within the same $0.5\,{\rm Mpc}$ aperture
for our clusters, finding that this measurement is feasible for 36 of
the 38 XXL clusters, and for all 10 COSMOS groups.  The best fit slope
parameter and intrinsic scatter for this fully self-consistent non-core excised relation are $b=1.61\pm0.12$, and
$\sigma_{(\ln\,M\,|\,T)}=0.42\pm0.06$. The mismatched aperture
uncertainty is therefore comparable to the statistical errors, and
does not alter our result.

\medskip

\noindent\emph{Selection function} -- The XXL-100-GC sample selection 
function needs to account for the flux-limit, survey volume, pointings and 
more. In the M-T relation this calculation is not trivial. We created a simplified 
toy model to test the bias in measured slope on a flux limited sample as a function 
of the correlation between X-ray luminosity and temperature. For this test we 
took a population of 10,000 groups and clusters with masses ($\rm 1\times 10^{13} < M_{500} < 1\times 10^{15}M_\odot$) and redshifts ($\rm 0<z<1.5$) from the \cite{Tinker2008} mass function. We converted the mass 
simultaneously to X-ray luminosity using the scaling relation in \cite{Maughan2014} and temperature 
using a relation of slope 1.5, normalisation 13.65. These were drawn from a bivariate Gaussian distribution with 
intrinsic scatter in log$_{10}$ of 0.4 and 0.3 for luminosity and temperature, respectively, and repeated for correlation coefficients
between luminosity and temperature from 0 to 1 in steps of 0.05. Each luminosity was then converted to a flux and 
a cut at $\rm 3 \times 10^{-14} ergs\ s^{-1}\ cm^{-2}$ was applied to replicate the selection on the XXL-100-GC sample. 
We drew 20 samples of 100 clusters before and after the flux cut for each of the correlation coefficients between L-T and 
fitted the mass-temperature relation for each of these samples. Comparing the bias between the scaling relation parameters measured before and after the flux cut as a function of the correlation between L-T shows a weak dependency. We expect the 
correlation coefficient between luminosity and temperature to be $\sim$0.3 \citep[e.g][]{Maughan2014}. In our model this 
corresponds to less than 5$\%$ bias in both slope and normalisation.  \cite{Kettula2015} apply a correction for Eddington bias to both masses and temperatures to a sample similar to ours in their scaling relation. Their results indicate a 10$\%$ bias on the slope when uncorrected for; however, this is detected at 0.7$\sigma$ significance. For the CCCP clusters used in this paper, a selection function model is not possible. The CCCP sample is selected from a variety of archived data and various selection criteria. We note that the selection function test above only applies to the XXL-only sample, but will be modelled comprehensively in a future XXL paper, when an alternative massive cluster sample with a well-defined selection function is available. 

\medskip

\noindent\emph{Outliers} -- One particular outlier in our sample is \texttt{XLSSC 110}. This system has been studied in detail by \citet{Verdugo2011} and is particularly interesting for the strong lensing features caused by a merger of three galaxies. For this system
the temperature is particularly low for the estimated mass. If we instead centre our shear profiles on the merger (corresponding to the BCG) we obtain a 25$\%$ higher mass. For this system the temperature may have been underestimated by the
exclusion of the AGN contaminated emission from the merger. \citet{Verdugo2011} use several methods to estimate the mass of this system but within a fixed radius.  Refitting the joint scaling relation excluding this system gives constraints of $ b=1.71\pm0.13$, $a=13.54\pm0.09$, and $\sigma_{\ln M|T}=0.41\pm0.06$.

\medskip

\noindent\emph{Mass bias on XXL-100-GC masses} -- To test the impact of biases on
the individually measured weak-lensing masses in the XXL sample on the masses derived from
the M--T relation, we perturbed the XXL masses down by increments of 10$\%$, refitted the joint M--T relation, and 
recomputed the masses of XXL-100-GC.  We find for offsets of 10, 20, and 30$\%$ in XXL masses, the 
resulting M--T derived masses, $M_{\rm 500,MT}$, will be lower by 0.04$\pm$0.02, 0.10$\pm$0.06, and 0.22$\pm$0.08, respectively.
Hence the systematics discussed in this section will have a relatively small influence on the XXL-100-GC masses computed from the M--T relation given the large uncertainties on the
linear regression parameters and temperature.

\subsection{Comparison with the literature}\label{sec:literature}

The mass-temperature relation fitted to the 96 clusters and groups
spanning $ T\simeq1-10\, {\rm keV}$ from XXL, COSMOS, and CCCP has a
slope of $b=1.67^{+0.14}_{-0.10}$.  This is 1.5$\sigma$ higher than the
self-similar prediction \citep{Kaiser1986}.  Most previous weak-lensing
based measurements of this relation have concentrated on higher
redshift samples, and/or a smaller (higher) temperature range \citep{Smith2005, Bardeau2007, Hoekstra2007, Okabe2010, Jee2011, Mahdavi2013}, thus precluding
useful comparison with our joint study of groups and clusters.
Our slope is marginally steeper (1.1$\sigma$ significance) than the most
comparable study, that of \cite{Kettula2013}, who obtained a slope
of $b=1.48^{+0.13}_{-0.09}$ for a sample of 65 groups and clusters
spanning a similar temperature and redshift range to ours.  The main
difference between their study and ours is that ours includes 38 new
systems from XXL-100-GC, we use the latest CCCP masses and the
temperatures are measured in different ways.  We measure temperatures
within a fixed metric aperture of $300\,{\rm kpc}$, whereas \citeauthor{Kettula2013} measure temperatures within an annulus that excludes the core
and scales with the mass of the cluster, $0.1r_{\rm 500,WL}<R<0.5r_{\rm 500,WL}$.
Nevertheless, within the current statistical precision the intercept and slope of the 
respective relations agree (Figure \ref{fig:compare}). We also note that the predicted 
self-similar slope applies to relations based on core-excised temperature
measurements.
We also express the normalisation of these two relations and those of
others from the literature as the mass of a cluster at $z=0.3$ with a
temperature of $T=3\, {\rm keV}$ to facilitate comparison between
relations that differ in the details of how they are defined. We see that 
the relations based on weak-lensing calibrated mass in the group regime
favour $\sim 40\%$ higher normalisations than hydrostatic relations at $\sim1-2 \sigma$. Although the bias correction applied by \cite{Kettula2015} can reproduce the self-similar slope, it has a negligible effect on the mass estimated at fixed T = 3 keV and z = 0.3 (Figure \ref{fig:compare})

Two of our clusters (\texttt{XLSSC 091} and \texttt{XLSSC 006})
also appear in \cite{Kettula2015} under their XID 111180 and 102760, using the same CFHTLenS survey data. The former has a spectroscopic redshift of 0.185\citep{Mirkazemi2015}, whereas the latter has a photometric measurement of 0.47\citep{Gozaliasl2014}, compared to our values of 0.186 and 0.429. For \texttt{XLSSC 091} and \texttt{XLSSC 006} respectively, the right ascension and declination are measured in XXL to be 37.926, -4.881 and 35.438, -3.772, whereas they appear in table 1 of \cite{Kettula2015} at 37.9269, -4.8814 and 35.4391, -3.7712. The respective offsets are $\sim$3.5" and $\sim$4.9".
They measure masses $ M_{\rm 500,WL} = 8.5\pm 2.1 \times10^{14}h_{70}^{-1} \rm M_\odot $ and $ 5.5\pm3.3 \times10^{14}h_{70}^{-1} \rm M_\odot$ and temperatures of $ T = 5 \pm 0.6 \,\rm keV$ and $\rm 8.2 \pm 5.6\, keV$. These agree with our masses and temperatures within the statistical errors.
 
Most studies of the mass-temperature relation of groups and clusters
have relied on X-ray data to estimate mass, and thus assumed that the
intracluster medium is in hydrostatic equilibrium \citep[e.g.][]{Finoguenov2001,Sun2009,Eckmiller2011,Lovisari2015}.  These authors obtained slopes of $b\simeq1.65-1.75$ with a
statistical uncertainty of $\sim0.05$.  
The \citeauthor{Kettula2013} core-excised weak-lensing relation is in tension with the
hydrostatic results at the 1-2 $\sigma$ level suggesting that the difference between
the lensing and X-ray based mass-temperature relations is mass dependent.
The slope of our weak-lensing-based non-core excised mass-temperature relation is, however, in agreement with the slope of the hydrostatic mass-temperature relations. 

Several observational and theoretical studies have found that
hydrostatic equilibrium may not be a valid assumption in the most
massive clusters \citep[e.g.][]{Nagai2007,Mahdavi2008,Mahdavi2013,Shaw2010,Zhang2010,Rasia2012,Israel2015}.  The assumption of hydrostatic equilibrium has
not yet been explored in great detail in galaxy groups,
i.e.\ $ T\ls3\,{\rm keV}$; however, \cite{Borgani2004} pointed out
that the steep slope of the hydrostatic mass-temperature relation of
groups is hard to reproduce with simulations. More recent papers of \cite{LeBrun2014, Pike2014, Planelles2014} show that the reproducibility of scaling relations is dependent on the physics included in the simulation. Simulations including baryonic processes are expected to bias  scaling relations from the self-similar prediction with a stronger effect on low-mass systems where the baryons are more important. The statistical precision of our results
is not sufficient to test whether the validity of hydrostatic equilibrium is a function of halo mass.

\section{Summary}\label{sec:Summary}

We have presented a study of the mass-temperature relation of galaxy
groups and clusters spanning $ T\simeq1-10\,{\rm keV}$, based on
weak-lensing mass measurements.  Our main analysis is based on the 38
systems drawn from the XXL 100 brightest cluster sample, that also
lie within the footprint of the CFHTLenS shear catalog.  Here we
summarise the main results of this paper:

\begin{itemize}

\item We measured individual weak-lensing masses of clusters within 
  XXL-100-GC with careful checks on systematics. In this
  mass ($M_{500}\sim10^{13}-10^{15}M_\odot$) and
  temperature range ($1\lesssim T\lesssim6\,{\rm keV}$) this is currently the
  largest sample of groups and poor clusters with weak-lensing masses available for studying the mass-temperature relation.\medskip

\item We used the masses to calibrate the mass-temperature relation
  down to the group and poor cluster mass scale. This relation has a
  slope of $1.78^{+0.37}_{-0.32}$.\medskip

\item We find that the scatter in our XXL-only mass-temperature
  relation is dominated by systems with significant offsets between their
  BCG and X-ray centroids.  This suggests that ongoing/recent merging
  activity may act to increase the scatter by affecting the accuracy
  of our weak-lensing mass measurements and/or by perturbing the
  temperature of the merging systems.  We will return to this issue
  when better quality data become available.\medskip

\item We increased the sample by incorporating 48 massive clusters from
  CCCP and 10 X-ray selected groups from COSMOS. This extended sample
  spans the temperature range $ T\simeq1-10\,{\rm keV}$.  The
  mass-temperature relation for this extended sample is steeper than the self-similar prediction, with a slope of
  $1.67^{+0.14}_{-0.10}$ and intrinsic scatter of $\sigma_{\ln
    M|T}=0.41$.  We used this relation to estimate the mass of each
  member of XXL-100-GC; these masses are
  available in \citetalias{Giles2016}.\medskip

\item The slope of our mass-temperature relation is in agreement with
  relations based on assuming hydrostatic equilibrium favouring a steeper slope
  than self-similar. Whilst insignificant given the current
  uncertainties, this result is in tension with previous weak-lensing studies 
  that suggest non-thermal pressure support being more significant in lower mass systems.  
  However, the offset in the normalisation of the relations estimated by comparing the mass of a $3\,{\rm keV}$ system
  at $z=0.3$ using the available relations implies that the hydrostatic mass of a 3 keV system is
  $\sim40\%$ lower than that obtained using a weak-lensing mass-temperature
  relation, which may indicate a halo mass dependent hydrostatic mass bias.

\end{itemize} 

Our future programme will extend mass-observable scaling relations for
groups and clusters in the XXL and related surveys to include other
mass proxies, including gas mass and $K$-band luminosity.  We will
also expand the sample of groups and poor clusters available for this
work as deeper weak-lensing data becomes available for XXL-N from
Hyper Suprime-CAM, and high-quality weak-lensing data become available
for XXL-S from our ongoing observations with Omegacam on the ESO VLT Survey Telescope.
These enlarged samples and the improved statistical precision will also
motivate careful modelling and the incorporation of the selection function
into our analysis.

\begin{acknowledgements}
This work is based on observations obtained with MegaPrime/MegaCam, a
joint project of CFHT and CEA/DAPNIA, at the Canada-France-Hawaii
Telescope (CFHT), which is operated by the National Research Council
(NRC) of Canada, the Institut National des Sciences de l'Univers of
the Centre National de la Recherche Scientifique (CNRS) of France, and
the University of Hawaii. This research used the facilities of the
Canadian Astronomy Data Centre operated by the National Research
Council of Canada with the support of the Canadian Space Agency.
CFHTLenS data processing was made possible thanks to significant
computing support from the NSERC Research Tools and Instruments grant
program.\\
We thank the anonymous referee for useful comments and Doug Applegate, Mark Birkinshaw, Gus Evrard, Will Farr,
Catherine Heymans, Henk Hoekstra, Kimo Kettula, Sarah Mulroy, Nobuhiro Okabe, Thomas
Reiprich, Tim Schrabback, Mauro Sereno and Trevor Sidery for helpful discussions
and assistance. ML acknowledges a Postgraduate
Studentship from the Science and Technology Facilities Council.  GPS
acknowledges support from the Royal Society. FP acknowledges support from the BMBF/DLR grant 50 OR 1117, the DFG grant RE 1462-6 and the DFG Transregio Programme TR33. GPS, FZ, TJP, BM, PG
acknowledge support from the Science and Technology Facilities
Council.
\end{acknowledgements}

\bibliographystyle{aa}
\bibliography{}
\onecolumn
\begin{appendix}
\section{Shear profiles}
\begin{figure}[!ht]
\vspace{-3cm}
\includegraphics[width=0.9\linewidth]{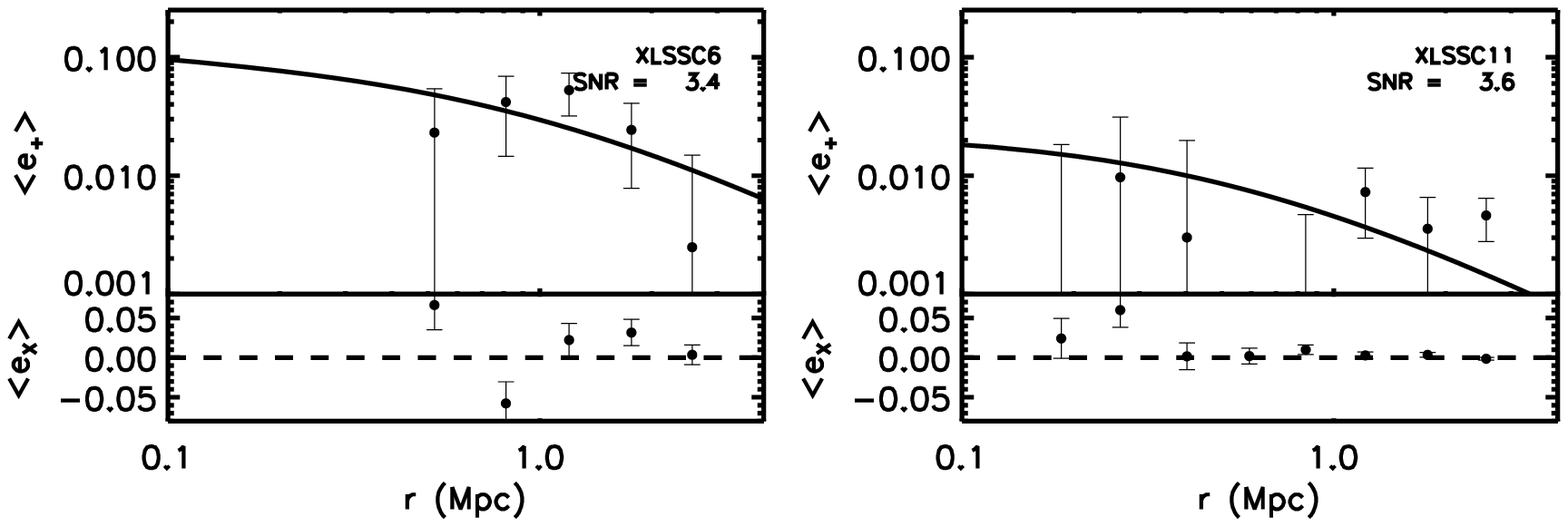}
\vspace{-6cm}

\includegraphics[width=0.9\linewidth]{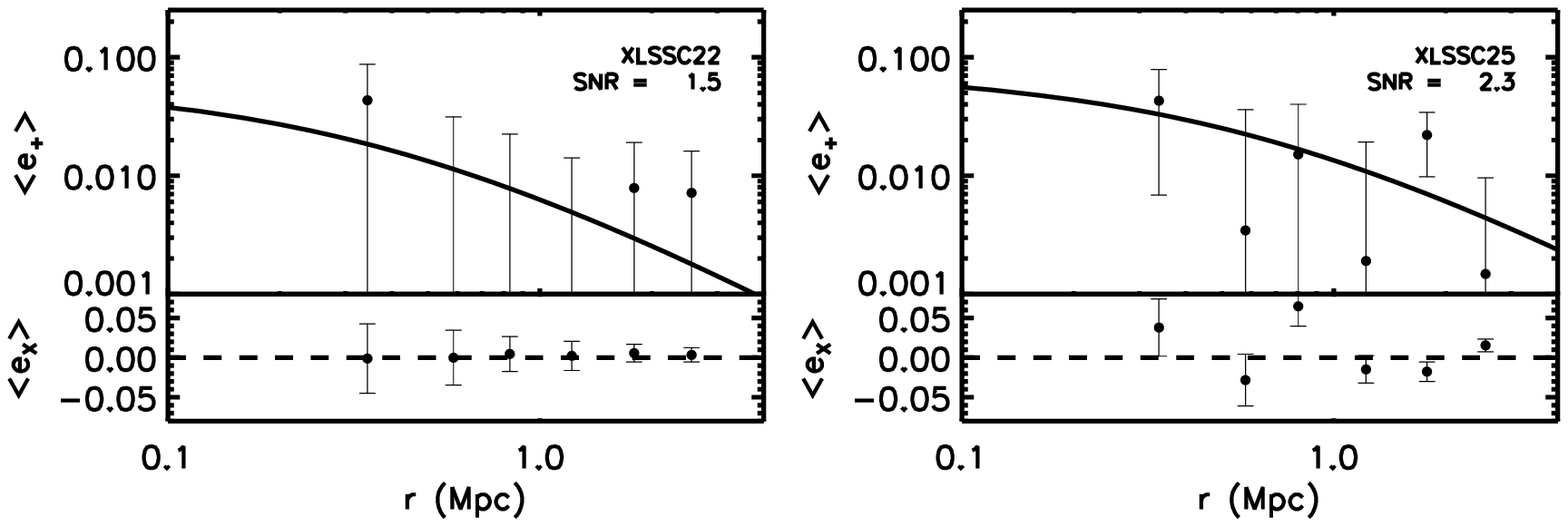}
\vspace{-6cm}

\includegraphics[width=0.9\linewidth]{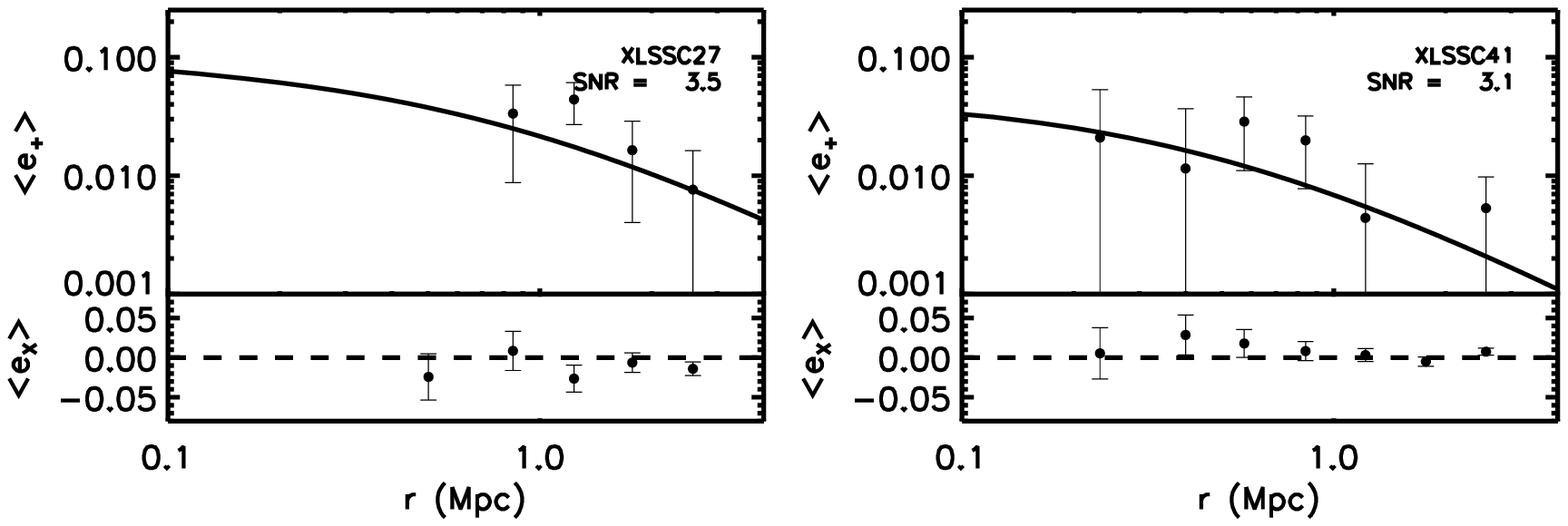}
\vspace{-6cm}

\includegraphics[width=0.9\linewidth]{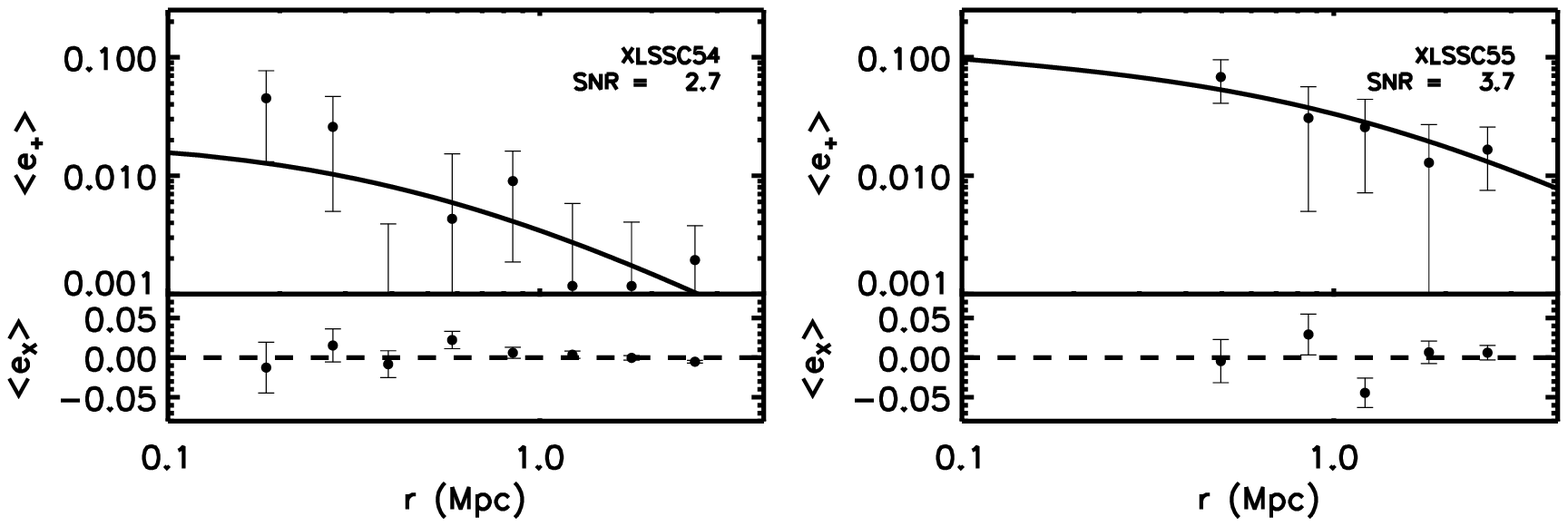}
\vspace{-4cm}
\caption{ Tangential and cross-component ellipticity as a function of
  distance from cluster centre. \label{fig:shprof1}}
\end{figure}

\begin{figure*}	
\vspace{-1cm}
\includegraphics[width=0.9\linewidth]{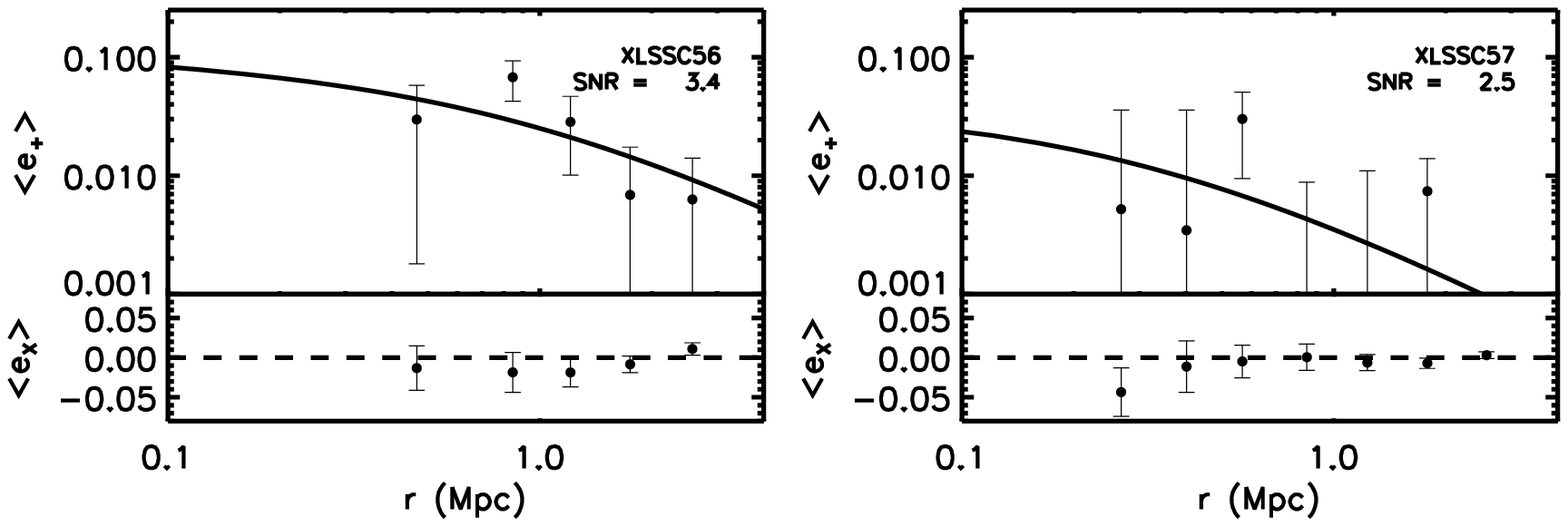}
\vspace{-6cm}

\includegraphics[width=0.9\linewidth]{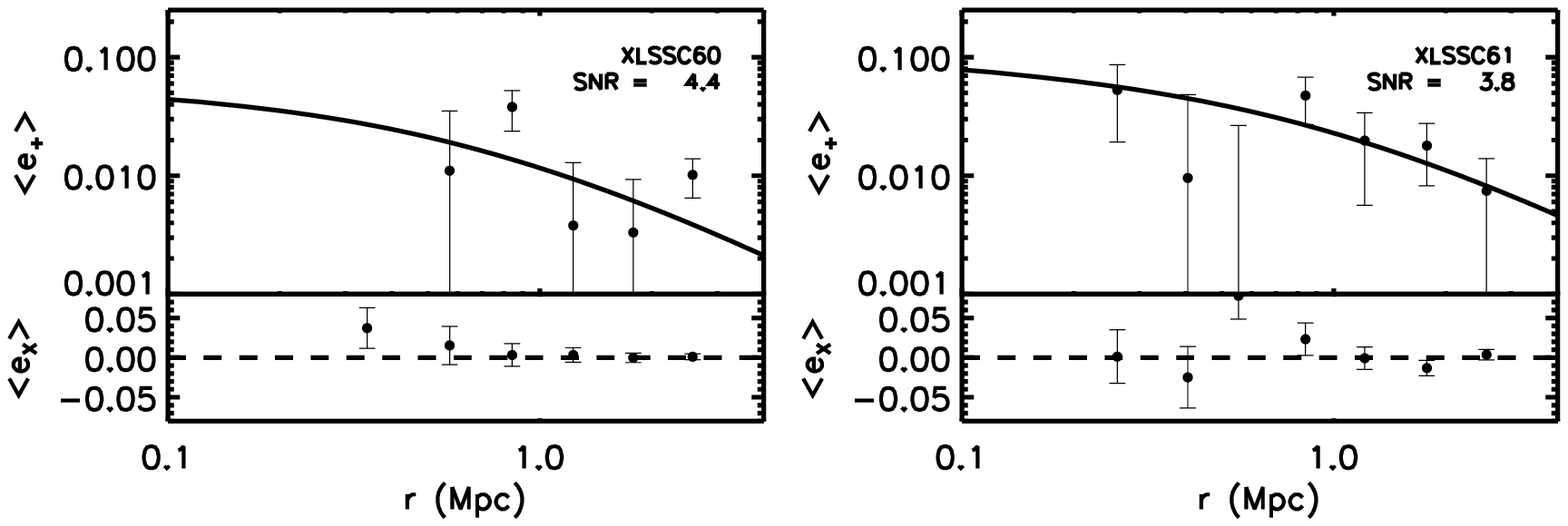}
\vspace{-6cm}

\includegraphics[width=0.9\linewidth]{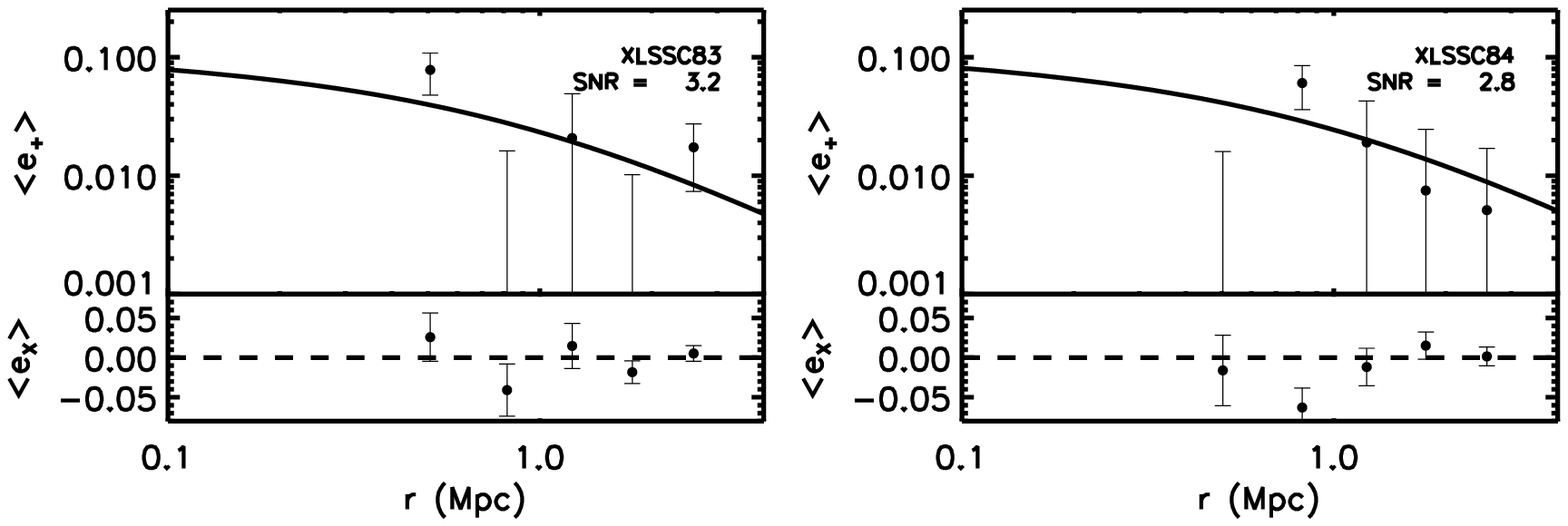}
\vspace{-6cm}

\includegraphics[width=0.9\linewidth]{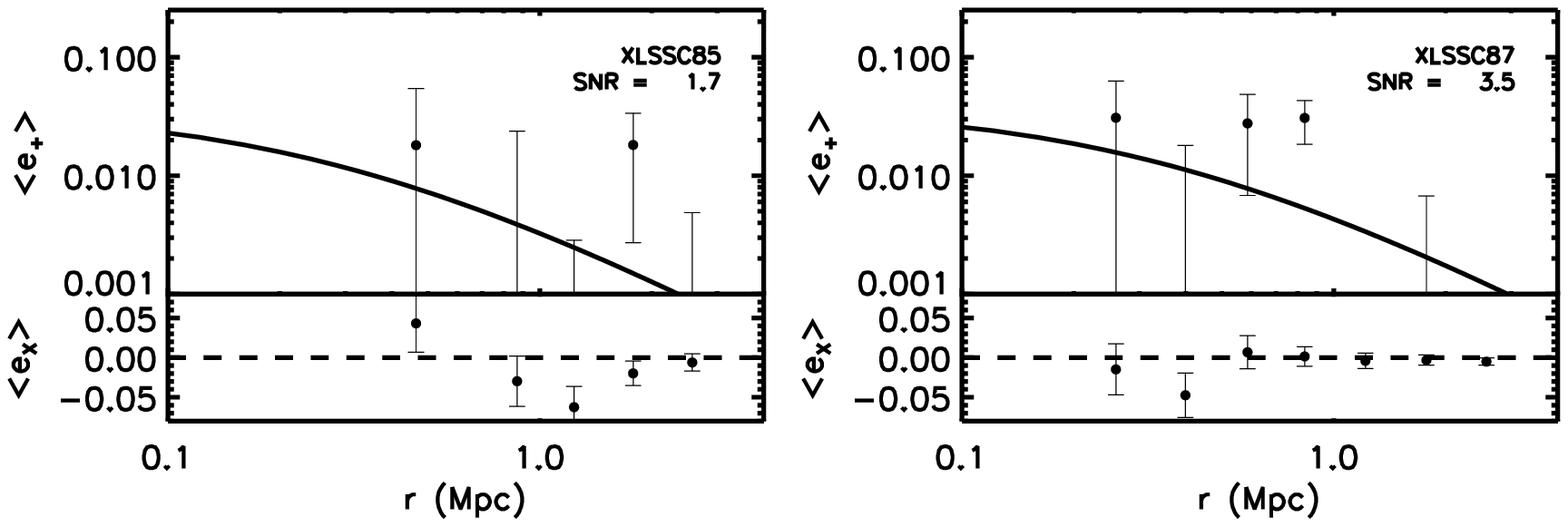}
\vspace{-4cm}
\caption{ Tangential and cross-component ellipticity as a function of
  distance from cluster centre. \label{fig:shprof2}}
\end{figure*}

\begin{figure*}	
\vspace{-1cm}
\includegraphics[width=0.9\linewidth]{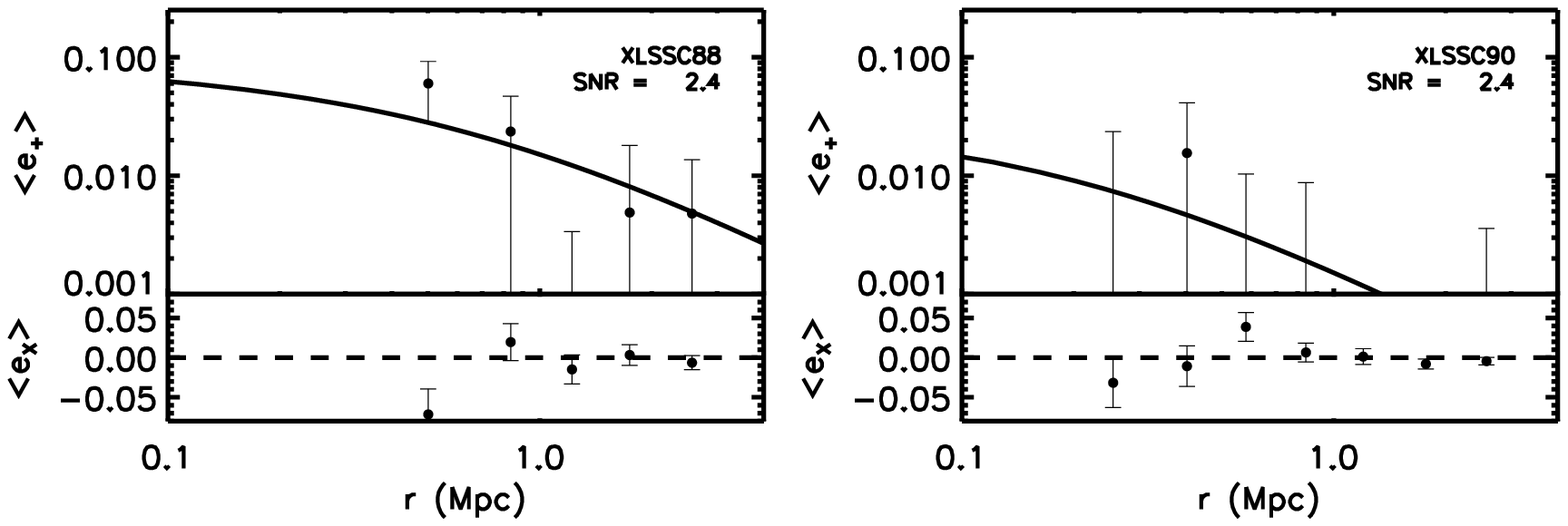}
\vspace{-6cm}

\includegraphics[width=0.9\linewidth]{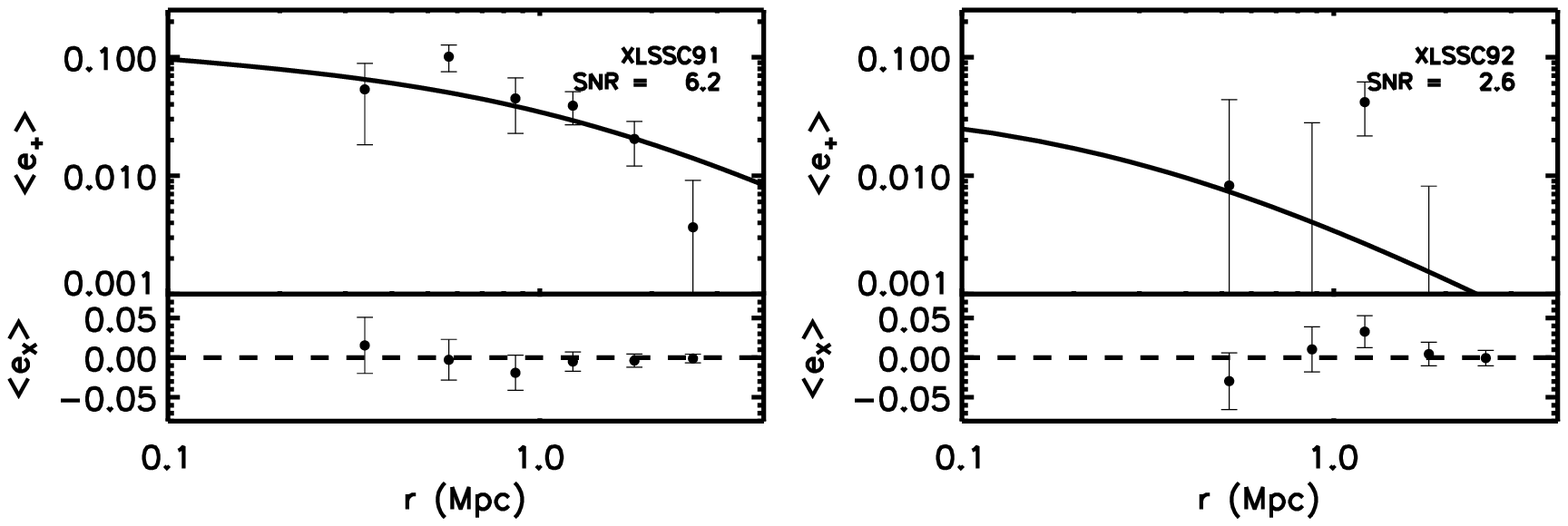}
\vspace{-6cm}

\includegraphics[width=0.9\linewidth]{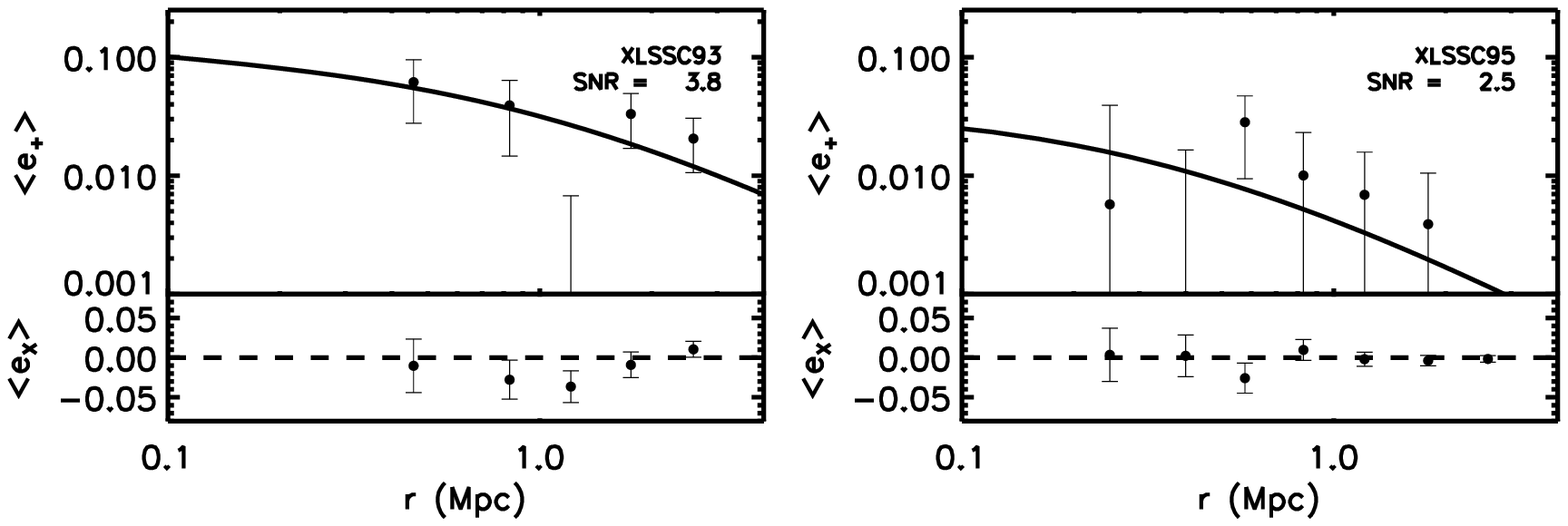}
\vspace{-6cm}

\includegraphics[width=0.9\linewidth]{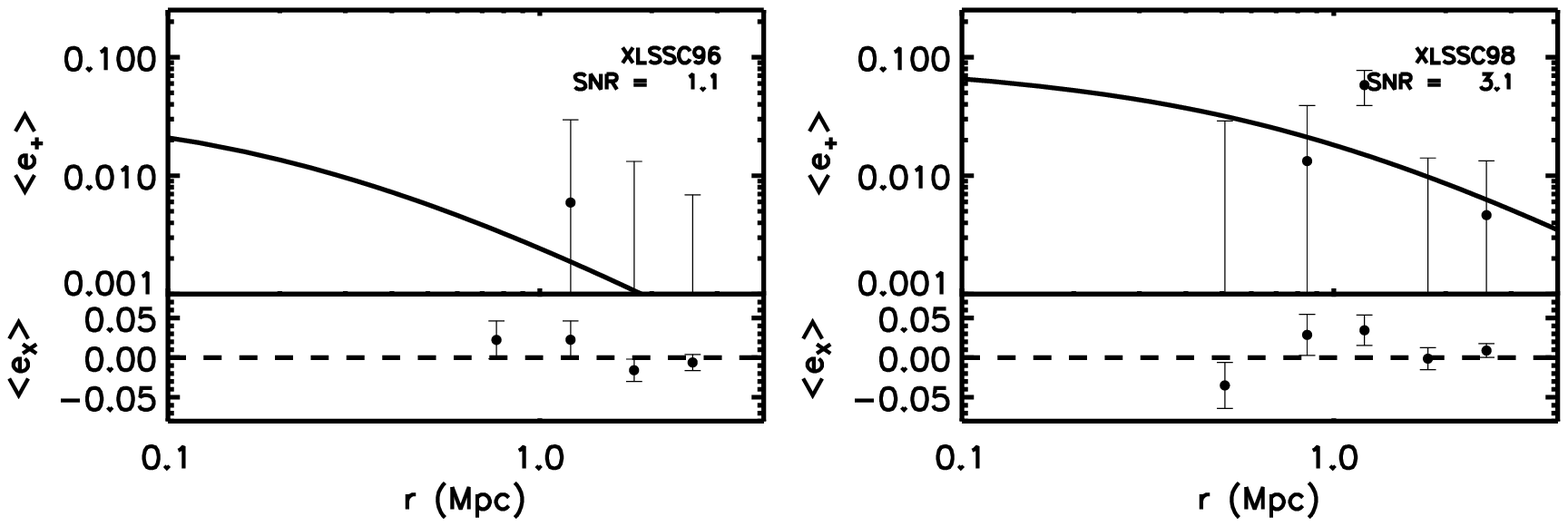}
\vspace{-4cm}
\caption{ Tangential and cross-component ellipticity as a function of
  distance from cluster centre. \label{fig:shprof3}}
\end{figure*}

\begin{figure*}	
\vspace{-1cm}
\includegraphics[width=0.9\linewidth]{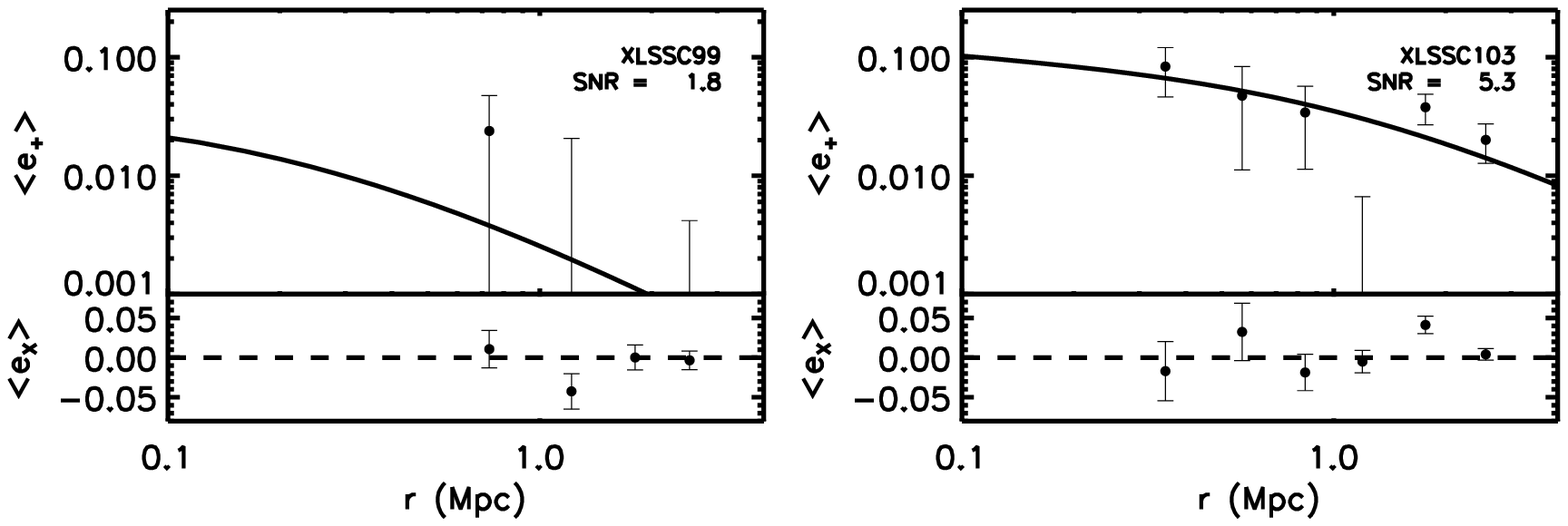}
\vspace{-6cm}

\includegraphics[width=0.9\linewidth]{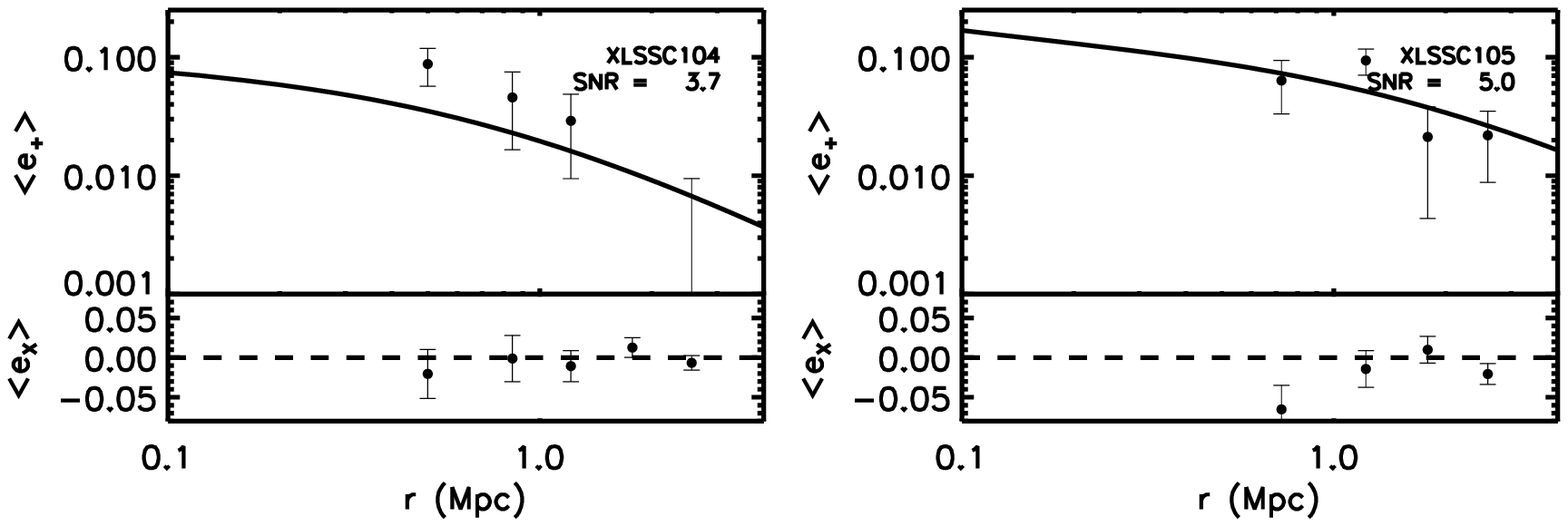}
\vspace{-6cm}

\includegraphics[width=0.9\linewidth]{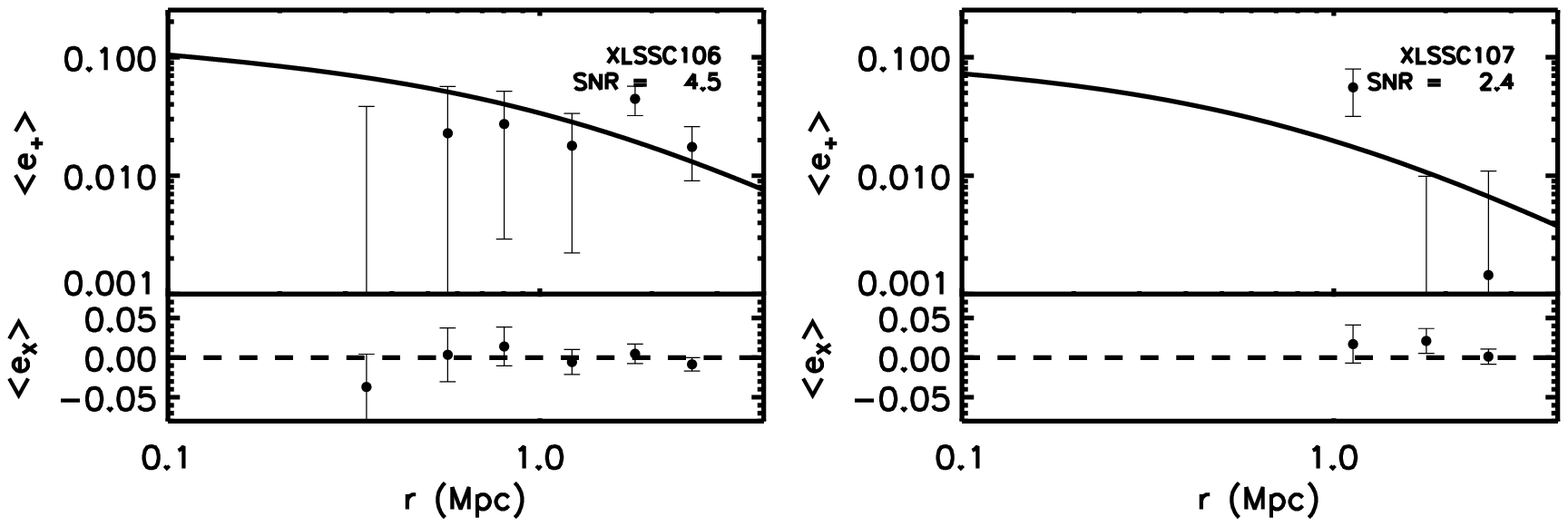}
\vspace{-6cm}

\includegraphics[width=0.9\linewidth]{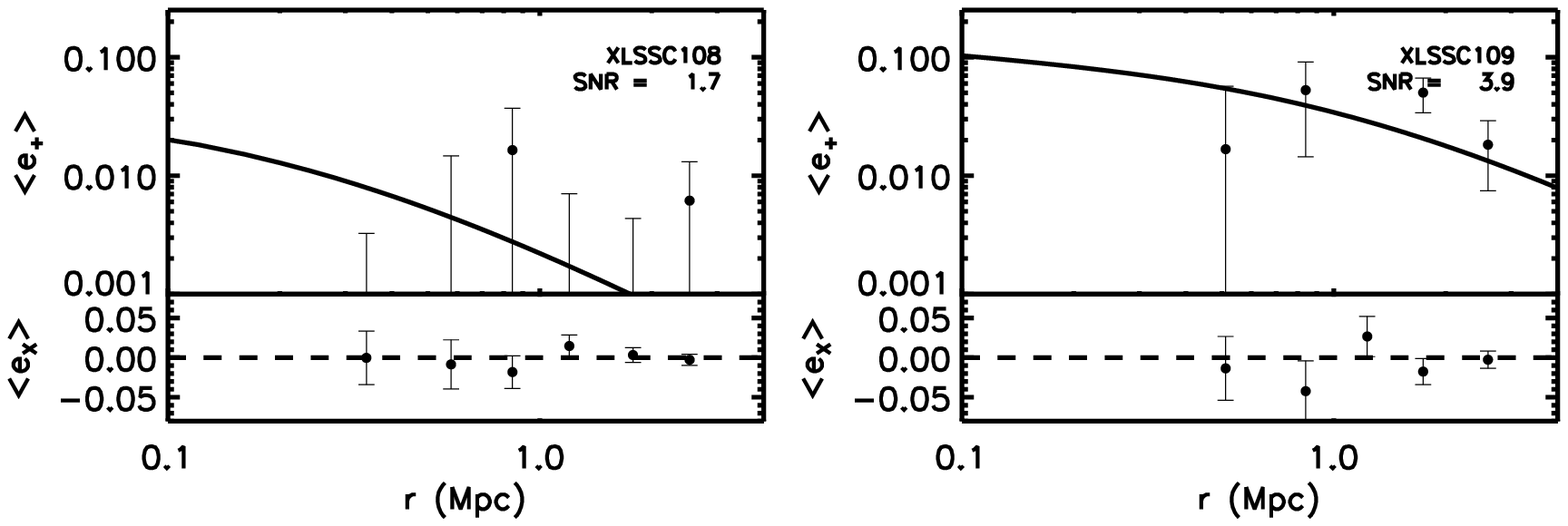}
\vspace{-4cm}
\caption{ Tangential and cross-component ellipticity as a function of
  distance from cluster centre. \label{fig:shprof4}}
\end{figure*}

\begin{figure*}	
\vspace{-1cm}
\includegraphics[width=0.9\linewidth]{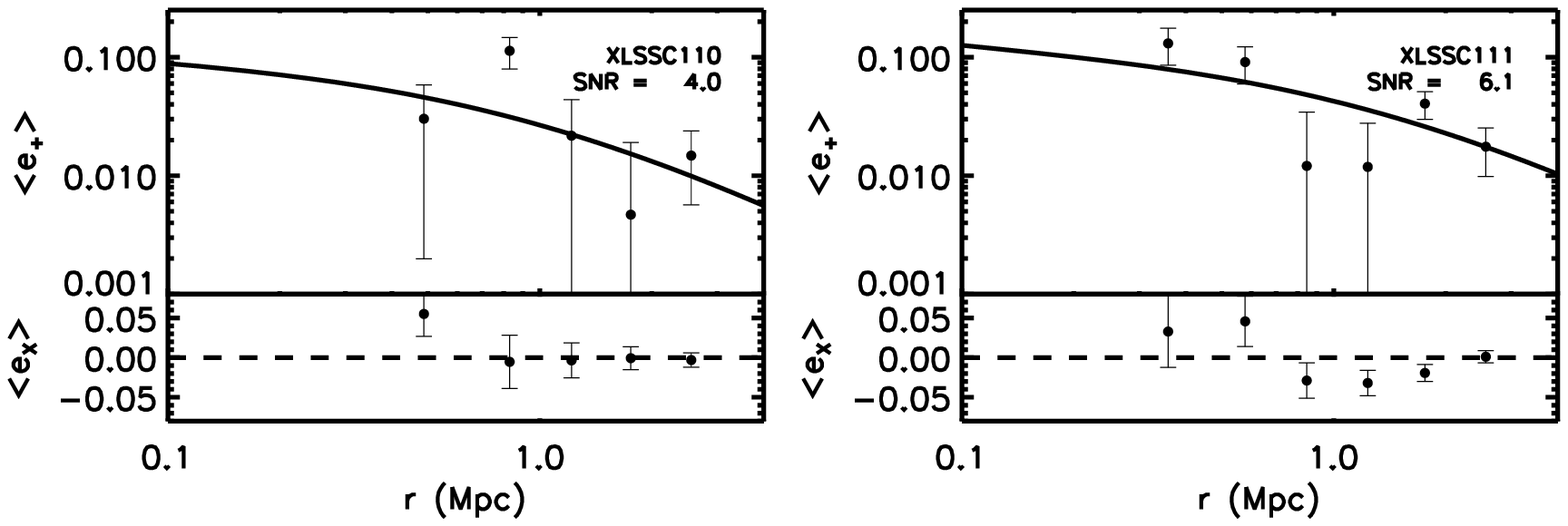}
\vspace{-6cm}

\includegraphics[width=0.9\linewidth]{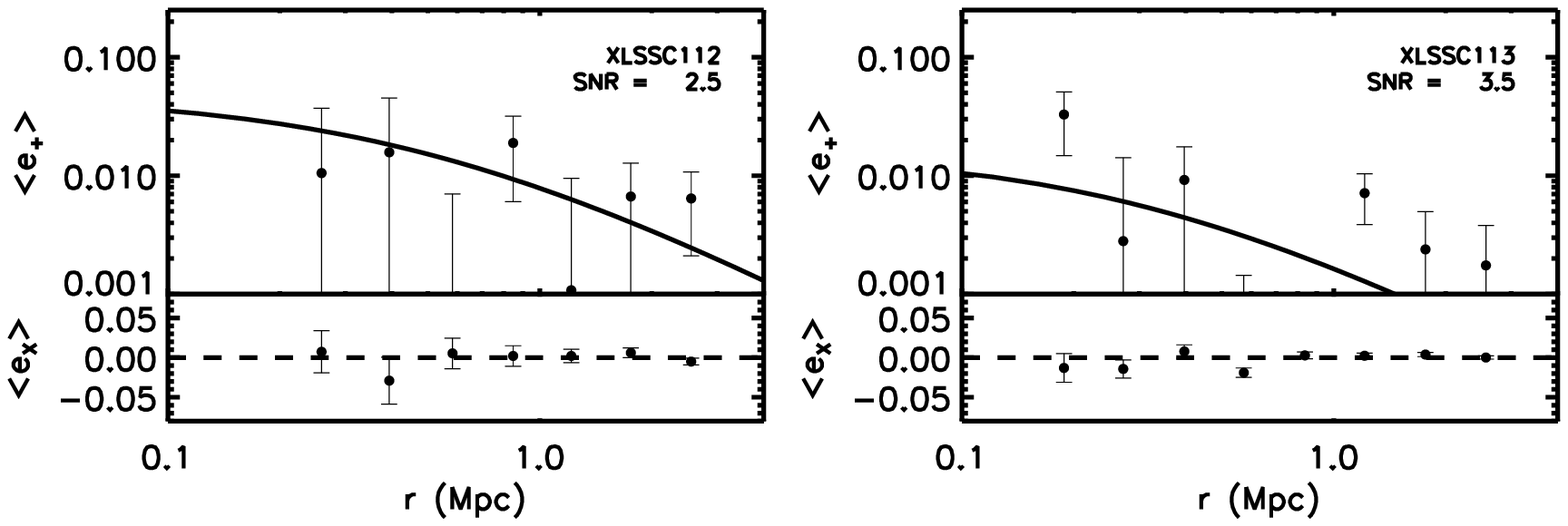}
\vspace{-6cm}

\includegraphics[width=0.9\linewidth]{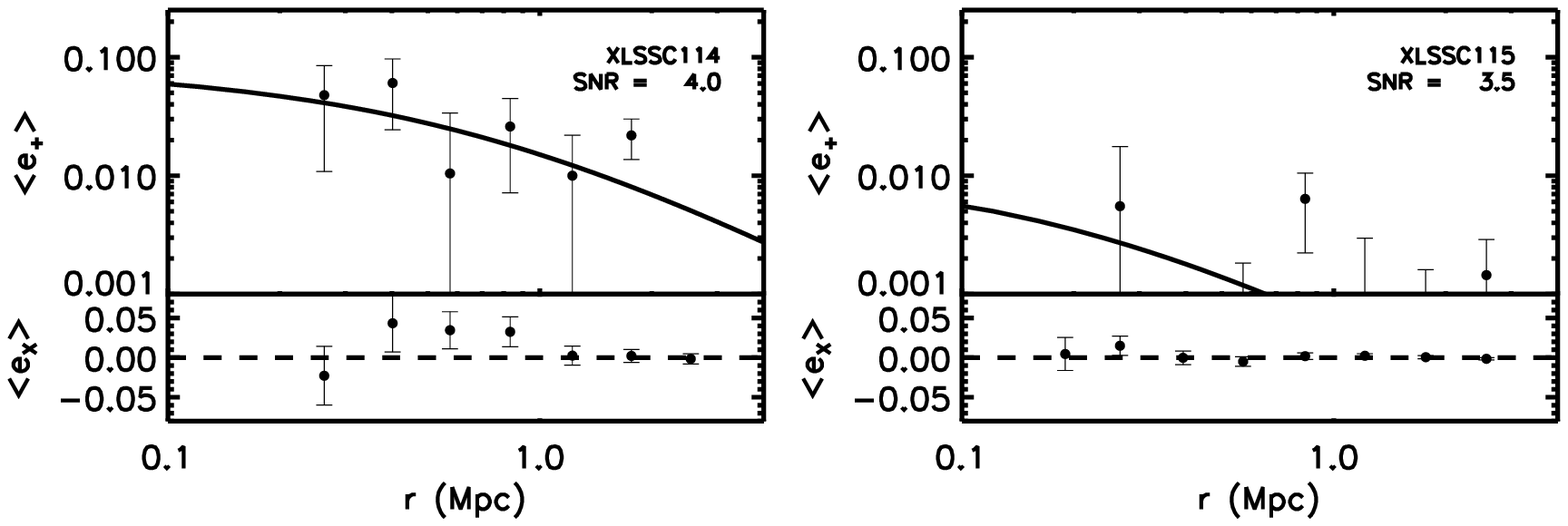}
\vspace{-4cm}
\caption{ Tangential and cross component ellipticity as a function of
  distance from cluster centre. \label{fig:shprof5}}
\end{figure*}

\end{appendix}

\end{document}